\documentclass[pdflatex,sn-nature]{sn-jnl}

\usepackage{graphicx}%
\usepackage{multirow}%
\usepackage{amsmath,amssymb,amsfonts}%
\usepackage{amsthm}%
\usepackage{mathrsfs}%
\usepackage[title]{appendix}%
\usepackage{xcolor}%
\usepackage{textcomp}%
\usepackage{manyfoot}%
\usepackage{booktabs}%
\usepackage{algorithm}%
\usepackage{algorithmicx}%
\usepackage{algpseudocode}%
\usepackage{listings}%

\begin{document}
\title{Particle Production and Density Fluctuations of Non-classical Inflaton in Coherent Squeezed Vacuum State of Flat FRW Universe}
\author[1]{\fnm{Dhwani} \sur{Gangal}}

\author[1]{\fnm{Sudhava} \sur{Yadav}}

\author*[1]{\fnm{K.K.} \sur{Venkataratnam}}\email{kvkamma.phy@mnit.ac.in}

\affil[1]{\orgdiv{Department of Physics}, \orgname{Malaviya National Institute of Technology}, \orgaddress{\street{J. L. N. Marg}, \city{Jaipur}, \postcode{302017}, \country{India}}}

\abstract{ We study non-classical inflaton, which is minimally coupled to the semiclassical gravity in FRW universe in Coherent Squeezed Vacuum State (CSVS). We determined Oscillatory phase of inflaton, power-law expansion, scale factor, density fluctuations, quantum fluctuations and particle production for CSVS. We obtained an estimated leading solution of scale factor in CSVS proportional to $t^{2/3}$ follow similar diversification as demonstrated by Semiclassical Einstein Equation (SCEE) of gravity in matter dominated universe. We also studied the validity of SCEE in CSVS. By determining the quantum fluctuation for CSVS validity of uncertainty relation for FRW Universe also computed. The results shows that Quantum fluctuations doesn't depend on coherent parameter $\Upsilon$ as uncertainty relation doesn't effected by the displacement of $\Upsilon$ in phase space. We study the production of particles in CSVS for oscillating massive inflaton in flat FRW universe.}


\maketitle

\section{Introduction}\label{S1}
Universe has started prompt expansion just after Big Bang as suggested by the theory of inflation\cite{kubik_origin_2022, moore_big_2014}. The big-bang model successfully explain evolution of universe but the model has few limitations in explaining the problems related with flatness, structure formation, monopole, singularity, horizon, homogeneity etc. of the universe. Further these problems were well addressed by inflationary theory \cite{guth1981cosmological}. There are multiple explanations are  available of above problems but the simplest explanation is an exponential expansion of universe. Further, potential energy of an inflaton that dominates in total energy of universe. As for the same potential energy, inflaton field deploy negative pressure leads very rapid expansion of the universe. As soon as the inflation got over, quasi-periodic motion has started in inflaton field whose immensely slowly decreased with time. The quasi-periodic nature of inflaton field produces various particles \cite{green_cosmological_2022}. The conversion of inflaton field and development in universe started rethermalization in universe \cite{albrecht_inflation_1994, albrecht_reheating_1982, kofman_reheating_1994, allahverdi_reheating_2010}. The transition is well parameterized and optimized by using various reheating parameters \cite{cook_reheating_2015, yadav_reheating_2023, yadav2024reheating} to plays significant role in understanding of standard matter production in universe. 

 As per the cosmological assumptions, Universe is considered as an isotropic and homogeneous. The universal representation is establish on the basis of Friedmann and Einstein Eqs. of field. In classical gravity, Friedmann equations \cite{mohajan_friedmann_2013, suresh_particle_2004} are valid even at an initial stages of universe. Subsequent analysis of universe show that, quantum properties and fluctuations of matter also play significant role in cosmology. Semiclassical Theory of Gravity (SCTG) establish the fact, that gravitational field is based on quantize matter field \cite{kim_one-parameter_1999, finelli_quantum_1999, geralico_novel_2004, padmanabhan_gravity_2005}. Even sometimes quantum gravity effects were considered to be negligible at an early stage in absence of proper hypothesis of quantum gravity. So to understand same, matter fields and gravity must present quantum mechanically, but in the absence of suitable consistent quantum theory that can describe gravity quantum mechanically, the problem become difficult. Using SCTG, where Friedmann’s equation of classical gravity and homogeneous field with Friedmann-Robertson-Walker (FRW) metric in background assumed their validity at an initial stage of universe in most of the inflationary scenarios. Thus, a comprehensive depiction of the early universe can be formulated by employing classical gravity and quantized matter field(s). It's noteworthy to highlight that quantized matter field(s) with fluctuations remain crucial, even when we assume that the effects of quantum gravity are negligible \cite{mahajan_particle_2008, lachieze-rey_cosmic_1995, ellis_cosmological_1998, carvalho_scalar_2004}. 

Few researchers in the field had used the quantum properties of inflaton with inflationary theories and SCTG \cite{bak1998quantum, kim1999thermal}. These inflationary theories are based on initial thermal condition, effective potential, quantum effect of inflaton \cite{guth1985quantum}, probability distribution \cite{habib1992stochastic} etc., to study quantum mechanical inflaton in stochastic inflationary scenarios \cite{linde1994big}. The complete cosmic evolution can be explained using semiclassical quantum gravity, starting from pre-inflation to inflation period in matter-dominated universe. These theories also propose that during the oscillatory phase, both quantum and classical inflaton mechanisms contribute to a similar power-law expansion.

 Quantum consideration of inflaton relies on SCTG within the framework of quantum optical considerations \cite{kennard_zur_1927, venkataratnam_particle_2004}. Kennard et., al. in their analysis of wave packet introduced the concept of squeezed states  \cite{bakke_geometric_2009}. The same can be integrated with the semi-classical Friedmann-Robertson-Walker (FRW) universe \cite{pedrosa_gaussian_2009, stoica_friedmann-lemaitre-robertson-walker_2016, hu_anisotropy_1978, dhayal_quantum_2020}. The anisotropic and expanding nature of FRW universe were further analyzed \cite{fischetti_quantum_1979, hartle_quantum_1979, hartle_quantum_1980, hartle_quantum_1981}. These studies show discrepancy in results of semiclassical gravity and classical gravity with similar power-law expansion for quantum as well as classical inflaton. The correction to expansion demonstrate oscillatory behaviour in semiclassical gravity, that doesn’t show oscillatory behaviour. The particle production within universe have been delineated for a coherently oscillating inflaton field within the Friedmann-Robertson-Walker (FRW) Universe framework \cite{anderson_effects_1983, campos_semiclassical_1994, geralico_novel_2004-1, pedrosa_exact_2007, lopes_gaussian_2009}.  These studies highlight the significant role of quantum phenomena in an inflationary theories. Recently, the study of quantum behaviour in cosmology \cite{gangal2024density, robertson1936kinematics, shaviv2011did, zel1971creation, bergstrom2006cosmology, ellis1999cosmological, kuo_semiclassical_1993, caves_quantum-mechanical_1981, matacz_coherent_1994, suresh_thermal_2001, suresh_squeezed_1998, suresh_nonclassical_2001, venkataratnam_nonclassical_2010,venkataratnam_density_2008, venkataratnam_oscillatory_2010, venkataratnam_behavior_2013, dhayal_quantum_2020-1, lachieze-rey_cosmological_1999,takahashi_thermo_1996, xu_quantum_2007} using non-classical state grabbing much attention \cite{koh_gravitational_2004, lopes_gaussian_2009-1, lachieze-rey_theoretical_1999, lopes_gaussian_2009-2, sinha_[no_2003, shaviv_did_2011, berger1978classical, berger1981scalar, grishchuk1990squeezed, brandenberger1992entropy, brandenberger1993entropy, kuo1993semiclassical, matacz1993quantum, albrecht1994inflation, gasperini1993quantum, hu1994squeezed}.

In this work, we examined the massive inflaton field, taking into account its minimal coupling with gravity within FRW universe framework. We are utilizing the Coherent Squeezed Vacuum State (CSVS) to describe the field \cite{nieto1997displaced, ellis1999deviation, penzias1965measurement, handley_curvature_2021}. CSVS has great applications in cosmology for explaining entropy enhancement \cite{gasperini1993quantum}, production of particles, gravitational wave detection \cite{savage1986inhibition} and inflationary scenario \cite{albrecht1982cosmology} etc. In the second section, we describe energy-momentum tensor, third section talks about the formulation of CSVS, in fourth section we have demonstrated the oscillatory phase of inflaton, power-law expansion and scale factor. Fifth section shows the expression for density fluctuations in CSVS. In the sixth section, our aim to validate the uncertainty relation in cosmology by computing quantum fluctuations within CSVS. Seventh section, deals with particle creation in CSVS formalism. In section eighth, we describe the outcome of our research work. 

\section{Energy-Momentum Tensor}

Modern cosmological models are constructed using classical gravity principles, derived from Einstein's field Eqs. within FRW metric. For these models, background metric is typically considered as classical and matter fields are approached from a quantum mechanical perspective. Such theoretical frameworks are commonly referred to as semiclassical theories. Further, the Einstein field Eqs. in semi-classical theory of gravity is (where $\hslash$=c=1 and {G}=$\frac{1}{\mathit{m}_{\mathit{p}}^2}$)

\begin{equation} \label{1.1}
\mathcal{E}_{\mu \nu }=\frac{8\pi }{\mathit{m}_{\mathit{p}}^2}\left\langle \mathcal{T}_{\mu \nu }\right\rangle.
\end{equation}

Here $\mathcal{T}_{\mu \nu }$ is energy-momentum tensor and \(\mathcal{E}_{\mu \nu }\) is the Einstein tensor. The quantum state, which satisfies the time-dependent Schrödinger equation, can be expressed as  

\begin{equation} \label{1.2}
\overset{\wedge }{\mathcal{H}}\psi=\mathit{i}\frac{ \partial }{\partial \mathit{t}}\psi,
\end{equation}
here \(\overset{\wedge }{\mathcal{H}}\) is Hamiltonian operator. FRW space-time with generalized variables \(\left(\mathit{r}_1,\mathit{r}_2,\right. \mathit{r}_3\),
\(\mathit{r}_4\)) can be written as

\begin{equation} \label{1.3}
\mathit{d}\mathit{s}^{2 }= -\mathit{d}\mathit{r}_4^2 + \mathcal{G}^{2 }(\mathbf{t})\left(\mathit{d}\mathit{r}_1^2 + \mathit{d}\mathit{r}_2^2 + \mathit{d}\mathit{r}_3^2\right),
\end{equation}
here $\mathcal{G}(\mathbf{t})$ is known as scale factor.\\

Lagrangian density $\mathfrak{L}$ is expressed as 

\begin{equation} \label{1.4}
\mathfrak{L} = -\frac{1}{2}\left(m^2\Phi  ^2+\mathfrak{g}^{\mu \nu }\partial _{\mu }\Phi  \partial _{\nu }\Phi\right)\sqrt{(-\mathfrak{g})}.
\end{equation}

Where $\Phi$ is scalar field, now considering the scalar field $\Phi$ is homogeneous. Using metric (\ref{1.3}), equation (\ref{1.4}) can be re-written as

\begin{equation} \label{1.5}
{\mathfrak{L}} =\frac{1}{2}{\mathcal{G}}^3(\mathbf{t})(\dot{\Phi}^2-m^2\Phi^2).
  \end{equation}
  
Using Eq. (\ref{1.4}), the K-G Eq. written as

\begin{equation} \label{1.6}
\ddot{\Phi  }+3\frac{\dot{\mathcal{G}} (\mathbf{t})}{\mathcal{G} (\mathbf{t})}\dot{\Phi  }+ m^2\Phi  =0,
\end{equation}

where \(\frac{\dot{\mathcal{G}}(\mathbf{t})}{\mathcal{G}(\mathbf{t})}\)=$\mathfrak{H}$ is the Hubble parameter and $\overset{\wedge}\Pi$ is the momentum conjugate to $\overset{\wedge}\Phi$ is 

\begin{equation} \label{1.7}
\overset{\wedge}\Pi = \frac{ \partial \mathcal{L}}{\partial \dot\Phi}.
\end{equation}

Using quantization condition, Hamiltonian for inflaton field, which behaves like a time-dependent harmonic oscillator in a suitable quantum state is  

\begin{equation}  \label{1.8}
\langle
:\overset{\wedge }{\mathcal{H}_m}:\rangle=\frac{1}{2\mathcal{G}^3 (\mathbf{t})}\left\langle :\overset{\wedge }\Pi ^2:\right\rangle {}+\frac{1}{2}\mathcal{G}^3(\mathbf{t})m^2\left\langle :\overset{\wedge }\Phi ^2:\right\rangle {},
\end{equation}

 where $ \left\langle :\overset{\wedge }\Pi ^2:\right\rangle {} $ and $ \left\langle :\overset{\wedge }\Phi ^2:\right\rangle {} $ are normal ordered expectation values of $\overset{\wedge }\Pi ^2$ and $\overset{\wedge }\Phi ^2$. The temporal part of energy-momentum tensor is 

\begin{equation} \label{1.9}
\mathcal{T}_{00}= \mathcal{G}^3 (\mathbf{t})\left(\frac{1}{2}\dot{\Phi }^2+\frac{1}{2}m^2\overset{\wedge }\Phi ^2\right).
\end{equation}

\section{Formation of Coherent Squeezed Vacuum State}

Coherent state can be described as

\begin{equation} \label{3.1}
|\varUpsilon \rangle=\mathfrak{D}(\Upsilon )|0\rangle,
\end{equation}

here $\mathfrak{D}(\varUpsilon)$ is displacement operator, can be represented as

\begin{equation} \label{3.3}
\mathfrak{D}(\Upsilon )=\exp (\Upsilon \overset{\wedge }{\mathit{e}} ^{\dagger }-\Upsilon ^*\overset{\wedge }{\mathit{e}}).
\end{equation}

Coherent squeezed vacuum state is

\begin{equation} \label{3.2}
|\varUpsilon ,\zeta \rangle =\overset{\wedge}W
(\rho ,\Psi )\mathfrak{D}(\varUpsilon )|0\rangle ,
\end{equation}

where squeezing operator \(\overset{\wedge}W(\rho ,\Psi )\) is

\begin{equation} \label{3.4}
\overset{\wedge}W(\rho ,\Psi )=\exp \frac{\rho }{2}\biggr(\overset{\wedge }{\mathit{e}}^2\exp (-\mathit{i}\Psi )-\overset{\wedge}{\mathit{e}}^{\dagger 2}\exp(\mathit{i}\Psi)\biggr) ,
\end{equation}

here squeezing parameter $\rho $ can take values between 0 $\leq $ $\rho $ $\leq $ $\infty $ and $\Psi $ representing the squeezing angle, can vary between -$\Pi $ and $\Pi $. \(\overset{\wedge}W(\rho ,\Psi )\) have properties like

\begin{equation} \label{3.5}
\overset{\wedge}W ^{\dagger }\overset{\wedge }{\mathit{e}}\overset{\wedge}W =\overset{\wedge }{\mathit{e}} \text{cosh$\rho $} - \overset{\wedge
}{\mathit{e}} ^{\dagger }\text{sinh$\rho $}\exp (\mathit{i}\Psi ),
\end{equation}

\begin{equation} \label{3.6}
\overset{\wedge}W ^{\dagger }\overset{\wedge }{\mathit{e}}^{\dagger }\overset{\wedge}W =\overset{\wedge }{\mathit{e}}^{\dagger }\text{cosh$\rho
$} - \overset{\wedge }{\mathit{e}}\text{sinh$\rho $}\exp (-\mathit{i}\Psi ).
\end{equation}

The annihilation \(\overset{\wedge }{\mathit{e}}\) and creation \(\overset{\wedge } {\mathit{e}} ^{\dagger }\) operator has following properties  

\begin{equation} \label{3.7}
\overset{\wedge }{\mathit{e}}|n,\mathbf{t},\Phi \rangle =\sqrt{n}|n-1,\mathbf{t},\Phi \rangle, \\
\end{equation}

\begin{equation} \label{3.8}
\overset{\wedge }{\mathit{e}} ^{\dagger }
|n,\mathbf{t},\Phi\rangle =\sqrt{n+1}|n+1,\mathbf{t},\Phi\rangle,
\end{equation}

\begin{equation} \label{3.9}
\overset{\wedge }{\mathit{e}} ^{\dagger }\overset{\wedge }{(\mathbf{t})\mathit{e}(\mathbf{t})}|n,\mathbf{t},\Phi\rangle =n|n,\mathbf{t},\Phi\rangle.
\end{equation}

These operators adhere to a specific commutation relation

\begin{equation} \label{3.10}
\left[\overset{\wedge }{\mathit{e}},\overset{\wedge }{\mathit{e}} ^{\dagger }\right]=1.
\end{equation}

The annihilation and creation operators can be calculated as

\begin{equation} \label{3.11}
\overset{\wedge }{\mathit{e}}(\mathbf{t})=\Phi ^*(\mathbf{t})\overset{\wedge }\Pi -\mathcal{G}^3 (\mathbf{t})\dot{\Phi }^*(\mathbf{t})\overset{\wedge }\Phi, 
\end{equation}

\begin{equation} \label{3.12}
\overset{\wedge }
{\mathit{e}} ^{\dagger }(\mathbf{t})=\Phi (\mathbf{t})\overset{\wedge }\Pi -\mathcal{G}^3 (\mathbf{t})\dot{\Phi }(\mathbf{t})\overset{\wedge }\Phi.
\end{equation}

Utilizing Eqs. (\ref{3.7}-\ref{3.12}), the operators $\overset{\wedge }\Phi $ and $\overset{\wedge }\Pi$ exhibit the following relations
\begin{equation} \label{3.13}
\overset{\wedge }\Phi =\frac{1}{i}\left(\Phi ^*\overset{\wedge }{\mathit{e}} ^{\dagger }-\Phi \overset{\wedge }{\mathit{e}} \right),\\
\end{equation}

\begin{equation} \label{3.14}
\overset{\wedge }\Phi ^2=\left(2\overset{\wedge }{\mathit{e}} ^{\dagger }\overset{\wedge }{\mathit{e}}+1\right)\Phi ^*\Phi-\left(\Phi \overset{\wedge }{\mathit{e}}\right)^2 -\left(\Phi ^*\overset{\wedge }{\mathit{e}}
^{\dagger }\right)^2,
\end{equation}

\begin{equation} \label{3.15}
\overset{\wedge }\Pi =\text{i$\mathcal{G}$}^3 (\mathbf{t})\left(\dot{\Phi }\overset{\wedge }{\mathit{e}} -\dot{\Phi }^*\overset{\wedge }{\mathit{e}} ^{\dagger
} \right),\\
\end{equation}

\begin{equation} \label{3.16}
\overset{\wedge }\Pi ^2=\mathcal{G}^3 (\mathbf{t})\left[\left(2\overset{\wedge }{\mathit{e}} ^{\dagger }\overset{\wedge }{\mathit{e}}+1\right)\dot{\Phi }^*\dot{\Phi
}-\left(\dot{\Phi }\overset{\wedge }{\mathit{e}}\right)^2-\left(\dot{\Phi }^*\overset{\wedge }{\mathit{e}} ^{\dagger }\right)^2\right].
\end{equation}

\(\overset{\wedge }{\mathit{e}}\) and \(\overset{\wedge }{\mathit{e}} ^{\dagger }\) combined with $\mathfrak{D}$($\Upsilon $) as given by Eq. (\ref{3.3}), to produce subsequent characteristics.

\begin{equation} \label{3.17}
\mathfrak{D}^{\dagger }\overset{\wedge }{\mathit{e}} ^{\dagger }\mathfrak{D}=\Upsilon ^*+\overset{\wedge }{\mathit{e}} ^{\dagger},
\end{equation}

\begin{equation} \label{3.18}
\mathfrak{D}^{\dagger }\overset{\wedge }{\mathit{e}} \mathfrak{D}=\Upsilon+\overset{\wedge }{\mathit{e}}.
\end{equation}

Applying the operator $\mathfrak{D}$($\Upsilon $) and $\overset{\wedge}W(\rho ,\Psi )$ to the vacuum state yields the CSVS
\begin{equation} \label{3.19}
|\Upsilon ,\zeta ,0\rangle =\mathfrak{D}(\Upsilon )\overset{\wedge}W(\rho ,\Psi )|0\rangle .
\end{equation}

\section{Power-law Expansion of SCEE and Oscillatory phase of inflaton for CSVS}

In the semi-classical theory of gravity, space-time is treated as classical with quantized matter field. We consider oscillatory phase for inflaton field that is minimally accompanied to a flat FRW metric, under classical gravity, Friedmann Eqs. is

\begin{equation} \label{4.2}
\left(\frac{\dot{\mathcal{G}}(\mathbf{t})}{\mathcal{G}(\mathbf{t})}\right)^2=\frac{8\pi }{3\mathit{m}_{\mathit{p}}^2}\frac{\mathcal{T}_{00}}{\mathcal{G}^3
(\mathbf{t})},
\end{equation}

where, \(\mathcal{T}_{00}\) is inflaton energy density given in Eq. (\ref{1.9}). In terms of \(\left\langle :\overset{\wedge }{\mathcal{H}_m}:\right\rangle\), Friedmann Eqs. is

\begin{equation} \label{4.3}
\left(\frac{\dot{\mathcal{G}}(\mathbf{t})}{\mathcal{G}(\mathbf{t})}\right)^2=\frac{8\pi }{3\mathit{m}_{\mathit{p}}^2}\frac{1}{\mathcal{G}^3 (\mathbf{t})}\langle
:\overset{\wedge }{\mathcal{H}_m}:\rangle.
\end{equation}

Here Eqs. (\ref{1.8}, \ref{3.7}-\ref{3.12}, \ref{4.3}) are used to compute the Hamiltonian in the semiclassical Friedmann Eqs as

\begin{equation} \label{4.4}
\left(\frac{\dot{\mathcal{G}}(\mathbf{t})}{\mathcal{G}(\mathbf{t})}\right)^2=\frac{8\pi }{3\mathit{m}_{\mathit{p}}^2}\left[\left(\frac{1}{2}+n\right)\left(\dot{\Phi }(\mathbf{t})\dot{\Phi
}^*(\mathbf{t})+m^2\Phi (\mathbf{t})\Phi ^*(\mathbf{t})\right)\right],
\end{equation}
where n is number parameter, \(\Phi ^*(\mathbf{t})\) and $\Phi $($\mathbf{t}$) follow Eq.(\ref{1.6}) and Wronskian identity for the same is given as 

\begin{equation} \label{4.5}
\mathcal{G}^3 (\mathbf{t})\biggr[\Phi (\mathbf{t})\dot{\Phi }^*(\mathbf{t})- \dot{\Phi }(\mathbf{t})\Phi ^*(\mathbf{t})\biggr]=\mathit{i}.
\end{equation}

Using above, we determine SCEE for CSVS, using Eqs. (\ref{3.4}-\ref{3.9}, \ref{3.16}), we get $ \left\langle :\overset{\wedge }\Pi ^2:\right\rangle {} $ for CSVS as 

\begin{align} \label{4.6}
\langle :\overset{\wedge }\Pi^2 :\rangle {}_{\text{CSVS}}=&2\mathcal{G}^6 (\mathbf{t})\biggl[\biggl(\frac{1}{2}sinh^2\rho +\frac{1}{2}+\varUpsilon^* \varUpsilon \biggr)\dot{\Phi}^*\dot{\Phi}\nonumber\\
&-\biggl(\frac{1}{2}\sinh\rho\cosh\rho -\frac{\varUpsilon^{*2}}{2}\biggr)\dot{\Phi}^{*2}\nonumber\\
&-\biggl(\frac{1}{2}\sinh\rho\cosh\rho -\frac{\varUpsilon^2}{2}\biggr)\dot{\Phi}^2\biggr].
\end{align}

Using Eqs. (\ref{3.4}-\ref{3.9}, \ref{3.14}) after computation $ \left\langle :\overset{\wedge }\Phi ^2:\right\rangle {} $ for CSVS is

\begin{align} \label{4.7}
\left\langle :\overset{\wedge }\Phi ^2:\right\rangle {}_{\text{CSVS}}=&2\biggr[\biggr(\frac{1}{2}\text{Sinh}^2\rho +\frac{1}{2}+\varUpsilon ^*\varUpsilon\biggr){\Phi }^*{\Phi }\nonumber\\
&-\biggr(\frac{1}{2}\text{Sinh$\rho $\text{Cosh$\rho $}}-\frac{\varUpsilon^{*2}}{2}\biggr){\Phi }^{*2}\nonumber\\
&-\biggr(\frac{1}{2}\text{Sinh$\rho $\text{Cosh$\rho $}}-\frac{\varUpsilon ^2}{2}\biggr){\Phi }^2\biggr].
\end{align}

Using Eqs.(\ref{1.8}, \ref{4.6}-\ref{4.7}), we got the Hamiltonian in CSVS as 

\begin{align} \label{4.8}
\left\langle :\overset{\wedge }{\mathcal{H}}:\right\rangle_{\text{CSVS}}=&\mathcal{G}^3 (\mathbf{t})\biggr[\biggr(\frac{1}{2}\text{Sinh}^2\rho +\frac{1}{2}+\varUpsilon ^*\varUpsilon\biggr)\biggr(\dot{\Phi }^*\dot{\Phi }+m^2{\Phi }^*{\Phi }\biggr)\nonumber\\
& -\biggr(\frac{1}{2}\text{Sinh$\rho $\text{Cosh$\rho $}}-\frac{\varUpsilon ^{*2}}{2}\biggr)\biggr(\dot{\Phi }^{*2}+m^2{\Phi }^{*2}\biggr)\nonumber\\
&-\biggr(\frac{1}{2}\text{Sinh$\rho $\text{Cosh$\rho $}}-\frac{\varUpsilon ^2}{2}\biggr)\biggr(\dot{\Phi^2}+m^2{\Phi^2}\biggr)\biggr].
\end{align}

Substituting Eqs.(\ref{4.8}) in Eqs.(\ref{4.3}), the semiclassical Einstein equation for CSVS is 

\begin{align} \label{4.9}
\left(\frac{\dot{\mathcal{G}}(\mathbf{t})}{\mathcal{G}(\mathbf{t})}\right)^2_{\text{CSVS}}=&\frac{8\pi }{3\mathit{m}_{\mathit{p}}^2}\biggr[\biggr(\frac{1}{2}\text{Sinh}^2\rho +\frac{1}{2}+\varUpsilon ^*\varUpsilon\biggr)\biggr(\dot{\Phi }^*\dot{\Phi }+m^2{\Phi }^*{\Phi }\biggr)\nonumber\\
& -\biggr(\frac{1}{2}\text{Sinh$\rho $\text{Cosh$\rho $}}-\frac{\varUpsilon ^{*2}}{2}\biggr)\biggr(\dot{\Phi }^{*2}+m^2{\Phi }^{*2}\biggr)\nonumber\\
&-\biggr(\frac{1}{2}\text{Sinh$\rho $\text{Cosh$\rho $}}-\frac{\varUpsilon ^2}{2}\biggr)\biggr(\dot{\Phi^2}+m^2{\Phi^2}\biggr)\biggr].
\end{align}

Next, we derive the expression for scale factor based on the equation (\ref{4.9}). To determine the normalization constant, we utilize the Wronskian and impose boundary conditions for two independent solutions, as following

\begin{equation} \label{4.10}
\Phi(\mathbf{t})={\psi(\mathbf{t})\over {\mathcal{G}}^{3/2}(\mathbf{t})},
\end{equation}

Substitute Eqs. (\ref{4.10}) in (\ref{1.6}) then we get 

\begin{equation} \label{4.11}
\ddot{\psi}(\mathbf{t})+\left(m^2-{3\over4}\left({\dot{\mathcal{G}}(\mathbf{t})\over
\mathcal{G}(\mathbf{t})}\right)^2-{3\over2}{\ddot{\mathcal{G}}(\mathbf{t})\over\mathcal{G}(\mathbf{t})}\right)\psi(\mathbf{t})=0.
\end{equation}

Here oscillatory region follows inequality
 
\begin{equation} \label{4.12}
m^2>{3\over4}\left({\dot{\mathcal{G}}(\mathbf{t})\over \mathcal{G}(\mathbf{t})}\right)^2+{3\over2}{\ddot{\mathcal{G}}(\mathbf{t})\over \mathcal{G}(\mathbf{t})},
\end{equation}

further oscillatory solution of inflaton is

\begin{equation} \label{4.13}
\psi(\mathbf{t})={\exp(-i\int \chi(\mathbf{t})d\mathbf{t})\over\sqrt{2\chi(\mathbf{t})}},
\end{equation}

where 

\begin{equation} \label{4.14}
\chi^2(\mathbf{t}) = m^2-{3\over4}\left({\dot{\mathcal{G}}(\mathbf{t})\over
\mathcal{G}(\mathbf{t})}\right)^2-{3\over2}{\ddot{\mathcal{G}}(\mathbf{t})\over \mathcal{G}(\mathbf{t})}
+{3\over4}\left({\dot{\chi}(\mathbf{t})\over \chi(\mathbf{t})}\right)^2-{1\over2}
{\ddot{\chi}(\mathbf{t})\over \chi(\mathbf{t})}.
\end{equation}

Using equation (\ref{4.13}) in Eq. (\ref{4.10}) we get 

\begin{equation} \label{4.15}
{\Phi}(\mathbf{t}) = \frac{1}{\mathcal{G}^{3/2}(\mathbf{t})}\frac{1}{\sqrt{2\chi(\mathbf{t})}} exp(-i\int \chi(\mathbf{t}) d\mathbf{t}), 
\end{equation}

\begin{equation} \label{4.16}
{\Phi}^*(\mathbf{t}) = \frac{1}{\mathcal{G}^{3/2}(\mathbf{t})}\frac{1}{\sqrt{2\chi(\mathbf{t})}} exp(i\int \chi(\mathbf{t}) d\mathbf{t}), \ 
\end{equation}

\begin{equation} \label{4.17}
\dot{\Phi}(\mathbf{t}) = \frac{exp(-i\int \chi(\mathbf{t}) d\mathbf{t})}{\mathcal{G}^{3/2}(\mathbf{t})\sqrt{2\chi(\mathbf{t})}} \biggr[-\frac{3}{2} \frac{\dot{\mathcal{G}(\mathbf{t})}}{\mathcal{G}(\mathbf{t})} - \frac{1}{2}\frac{\dot{\chi(\mathbf{t})}}{\chi(\mathbf{t})} - i \chi(\mathbf{t}) \biggr], 
\end{equation}

\begin{equation} \label{4.18}
\dot{\Phi}^*(\mathbf{t}) = \frac{exp(i\int \chi(\mathbf{t}) d\mathbf{t})}{\mathcal{G}^{3/2}(\mathbf{t})\sqrt{2\chi(\mathbf{t})}} \biggr[-\frac{3}{2} \frac{\dot{\mathcal{G}(\mathbf{t})}}{\mathcal{G}(\mathbf{t})} -\frac{1}{2} \frac{\dot{\chi(\mathbf{t})}}{\chi(\mathbf{t})} + i \chi(\mathbf{t}) \biggr]. 
\end{equation}

The scale factor for CSVS can be obtained using Eqs. (\ref{4.15}-\ref{4.18}) in equation (\ref{4.9}) as

\begin{align} \label{4.19}
  \biggr(\frac{\dot{\mathcal{G}}(\mathbf{t})}{\mathcal{G}(\mathbf{t})}\biggr)^2_{CSVS}  &= \frac{4\pi}{3 m_p^2 \chi(\mathbf{t}) \mathcal{G}^3(\mathbf{t})} \biggr[ (sinh^2\rho  + \Upsilon ^*\Upsilon + \frac{1}{2}) \biggr(\frac{9}{4}\biggr(\frac{\dot{\mathcal{G}}(\mathbf{t})}{\mathcal{G}(\mathbf{t})}\biggr)^2 + \frac{1}{4}\biggr(\frac{\dot{\chi}(\mathbf{t})}{\chi(\mathbf{t})}\biggr)^2\nonumber\\ &+\frac{3}{2}\frac{\dot{\mathcal{G}}(\mathbf{t})}{\mathcal{G}(\mathbf{t})}\frac{\dot{\chi}(\mathbf{t})}{\chi(\mathbf{t})} + \chi^2(\mathbf{t}) + m^2\biggr) +(\frac{\sinh 2\rho}{4}e^{-i\varphi} - \frac{\Upsilon^{*2}}{2})exp(2i\int\chi(\mathbf{t})dt) \nonumber\\ & 
\times \biggr(\frac{9}{4}\biggr(\frac{\dot{\mathcal{G}}(\mathbf{t})}{\mathcal{G}(\mathbf{t})}\biggr)^2 + \frac{1}{4}\biggr(\frac{\dot{\chi}(\mathbf{t})}{\chi(\mathbf{t})}\biggr)^2+\frac{3}{2}\frac{\dot{\mathcal{G}}(\mathbf{t})}{\mathcal{G}(\mathbf{t})}\frac{\dot{\chi}(\mathbf{t})}{\chi(\mathbf{t})} -\chi^2(\mathbf{t}) \nonumber\\ &- i\chi(\mathbf{t})\biggr(\frac{3\dot{\mathcal{G}(\mathbf{t})}}{\mathcal{G}(\mathbf{t})} + \frac{\dot{\chi(\mathbf{t})}}{\chi(\mathbf{t})} \biggr)+m^2 \biggr) +(\frac{\sinh 2\rho}{4}e^{i\varphi} - \frac{\Upsilon^2}{2})exp(2i\int\chi(\mathbf{t})dt) \nonumber\\ & 
 \times \biggr(\frac{9}{4}\biggr(\frac{\dot{\mathcal{G}}(\mathbf{t})}{\mathcal{G}(\mathbf{t})}\biggr)^2 + \frac{1}{4}\biggr(\frac{\dot{\chi}(\mathbf{t})}{\chi(\mathbf{t})}\biggr)^2+\frac{3}{2}\frac{\dot{\mathcal{G}}(\mathbf{t})}{\mathcal{G}(\mathbf{t})}\frac{\dot{\chi}(\mathbf{t})}{\chi(\mathbf{t})} -\chi^2(\mathbf{t}) \nonumber\\ & - i\chi(\mathbf{t})\biggr(\frac{3\dot{\mathcal{G}(\mathbf{t})}}{\mathcal{G}(\mathbf{t})} + \frac{\dot{\chi(\mathbf{t})}}{\chi(\mathbf{t})} \biggr) +m^2 \biggr) \biggr],
  \end{align}
  
simplify the above equation as

\begin{align} \label{4.20}
  \mathcal{G}^3(\mathbf{t})_{CSVS}  &= \frac{4\pi}{3 m_p^2 \chi(\mathbf{t}) (\frac{\dot{\mathcal{G}}(\mathbf{t})}{\mathcal{G}(\mathbf{t})})^2}\biggr[ (sinh^2 \rho  + \Upsilon ^*\Upsilon + \frac{1}{2}) \biggr(\frac{9}{4}\biggr(\frac{\dot{\mathcal{G}}(\mathbf{t})}{\mathcal{G}(\mathbf{t})}\biggr)^2 + \frac{1}{4}\biggr(\frac{\dot{\chi}(\mathbf{t})}{\chi(\mathbf{t})}\biggr)^2\nonumber\\ &+\frac{3}{2}\frac{\dot{\mathcal{G}}(\mathbf{t})}{\mathcal{G}(\mathbf{t})}\frac{\dot{\chi}(\mathbf{t})}{\chi(\mathbf{t})} + \chi^2(\mathbf{t}) + m^2\biggr) +(\frac{\sinh 2\rho}{4}e^{-i\varphi} - \frac{\Upsilon^{*2}}{2})exp(2i\int\chi(\mathbf{t})dt) \nonumber\\ & 
\times \biggr(\frac{9}{4}\biggr(\frac{\dot{\mathcal{G}}(\mathbf{t})}{\mathcal{G}(\mathbf{t})}\biggr)^2 + \frac{1}{4}\biggr(\frac{\dot{\chi}(\mathbf{t})}{\chi(\mathbf{t})}\biggr)^2+\frac{3}{2}\frac{\dot{\mathcal{G}}(\mathbf{t})}{\mathcal{G}(\mathbf{t})}\frac{\dot{\chi}(\mathbf{t})}{\chi(\mathbf{t})} -\chi^2(\mathbf{t}) \nonumber\\ &- i\chi(\mathbf{t})\biggr(\frac{3\dot{\mathcal{G}(\mathbf{t})}}{\mathcal{G}(\mathbf{t})} + \frac{\dot{\chi(\mathbf{t})}}{\chi(\mathbf{t})} \biggr)+m^2 \biggr) +(\frac{\sinh 2\rho}{4}e^{i\varphi} - \frac{\Upsilon^2}{2})exp(2i\int\chi(\mathbf{t})dt) \nonumber\\ & 
 \times \biggr(\frac{9}{4}\biggr(\frac{\dot{\mathcal{G}}(\mathbf{t})}{\mathcal{G}(\mathbf{t})}\biggr)^2 + \frac{1}{4}\biggr(\frac{\dot{\chi}(\mathbf{t})}{\chi(\mathbf{t})}\biggr)^2+\frac{3}{2}\frac{\dot{\mathcal{G}}(\mathbf{t})}{\mathcal{G}(\mathbf{t})}\frac{\dot{\chi}(\mathbf{t})}{\chi(\mathbf{t})} -\chi^2(\mathbf{t}) \nonumber\\ & - i\chi(\mathbf{t})\biggr(\frac{3\dot{\mathcal{G}(\mathbf{t})}}{\mathcal{G}(\mathbf{t})} + \frac{\dot{\chi(\mathbf{t})}}{\chi(\mathbf{t})} \biggr) +m^2 \biggr) \biggr] ,
  \end{align}

  further using perturbation method as well as following approximation ansatzs as  
  \begin{align} \label{4.21}
\chi_0(\mathbf{t}) &=m
\end{align}

\begin{align} \label{4.22}
\mathcal{G}_0(\mathbf{t})&=\mathcal{G}_0t^{2/3},
\end{align}
  
to solve equation (\ref{4.20}), we use $\Upsilon = |\Upsilon| e^{i\theta}$ and  approximation ansatz (\ref{4.21}-\ref{4.22}), in equation (\ref{4.20}), hence we get next order solution as

\begin{align} \label{4.23}
\mathcal{G}_1(\mathbf{t})_{CSVS} &= \biggr[\frac{6\pi}{m_p^2}(\sinh^2 \rho  + \Upsilon ^*\Upsilon + \frac{1}{2}) m\mathbf{t}^2 \biggr(1 + \frac{1}{2 m^2 \mathbf{t}^2}\biggr) \nonumber\\ & +\frac{3\pi \mathbf{t}^2}{2m_p^2} \sinh2\rho \biggr(\frac{\cos(\varphi-2m\mathbf{t})}{m\mathbf{t}^2}-\frac{2}{\mathbf{t}}\sin(\varphi-2m\mathbf{t})\biggr)
 \nonumber\\ & 
 - \frac{3\pi \Upsilon^2 \mathbf{t}^2}{m_p^2}  \biggr(\frac{\cos2(\theta-m\mathbf{t})}{m\mathbf{t}^2}-\frac{2}{\mathbf{t}}\sin2(\theta-m\mathbf{t})\biggr)
 \biggr]^\frac{1}{3}.
\end{align}

The scale factor of CSVS is shown in equation (\ref{4.23}), is a function of squeezing parameter $\rho$, coherent angle $\theta$, angle of squeezing $\varphi$ and coherent state parameter $\Upsilon$. At resonance condition when we set limit $m\mathbf{t} >> 1$, $\theta = m\mathbf{t}$, $\varphi = 2m\mathbf{t}$ and $\rho = 0$, the scale factor  $\mathcal{G}_1(\mathbf{t})_{CSVS}$ directly proportional to $t^{2/3}$, i.e., the study reveled that the semi-classical Einstein equation of gravity gives similar power law expansion of scale factor as classical Einstein equation of gravity. 

\section{Validity of SCTG and Density fluctuations in CSVS}

The Validity of SCTG can be obtained by determining density fluctuation using SCEE. Following connection are used to study the inflaton for various quantum states in FRW universe. 

\begin{equation} \label{5.1}
\triangle = \langle :T_{\mu \nu }^2:\rangle  - \langle :T_{\mu \nu }:\rangle ^2.
\end{equation}

Here $T_{\mu \nu }$ is energy momentum tensor. The density fluctuations $\triangle$ for CSVS is computed using Eq. (\ref{5.1}), where \(\left\langle :T_{00}^2:\right\rangle {}_{\text{CSVS}}\) is given as

\begin{align} \label{5.2}
\left\langle :T_{00}^2:\right\rangle_{\text{CSVS}}=
&\frac{1}{4}\mathcal{G}^6(\mathbf{t})m^{4}\left\langle :\overset{\wedge }\Phi ^4:\right\rangle_{\text{CSVS}}+\frac{m^{2
}}{4}\left\langle:\overset{\wedge }\Phi ^2\overset{\wedge }\Pi ^2:\right\rangle _{\text{CSVS}}\nonumber\\
&+\frac{m^{2 }}{4}\left\langle :\overset{\wedge }\Pi ^2\overset{\wedge }\Phi ^2:\right\rangle_{\text{CSVS}}+\frac{1}{4\mathcal{G}^6
(\mathbf{t})}\left\langle :\overset{\wedge }\Pi ^4:\right\rangle_{\text{CSVS}}.
\end{align}

Using Eqs. (\ref{3.5}-\ref{3.6}, \ref{3.11}-\ref{3.19}) values of \(\left\langle :\overset{\wedge }\Pi ^4:\right\rangle {}_{\text{CSVS}}\) is  

\begin{align} \label{5.3}
\left\langle :\overset{\wedge }\Pi ^4:\right\rangle_{\text{CSVS}}=
&\frac{\mathcal{G}^6(\mathbf{t})}{4m^{2 }\mathbf{t}^4}\biggr[3+\Upsilon ^{*4}+\Upsilon^4+6\Upsilon
^{*2}\Upsilon ^2+12\Upsilon ^*\Upsilon -6\Upsilon ^{*2}-6\Upsilon ^2\nonumber\\
&-4\Upsilon ^{*3}\Upsilon -4\Upsilon ^*\Upsilon ^3+ 12\text{Sinh}^4\rho+12\text{Cosh}^2\rho\text{Sinh}^2\rho+24\text{Cosh$\rho
$} \text{Sinh}^3\rho \nonumber\\
&+12\text{Cosh$\rho $} \text{Sinh$\rho $}\left\{1+2\Upsilon ^*\Upsilon -\Upsilon ^{*2}-\Upsilon
^2\right\}\nonumber\\
&+12\text{Sinh}^2\rho\left\{1+2\Upsilon ^*\Upsilon -\Upsilon^{*2}-\Upsilon ^2\right\}\biggr].
\end{align}

Using Eqs. (\ref{3.5}-\ref{3.6}, \ref{3.11}-\ref{3.19}) values of \(\left\langle :\overset{\wedge }\Phi ^4:\right\rangle _{\text{CSVS}}\) is 

\begin{align} \label{5.4}
\left\langle :\overset{\wedge }\Phi ^4:\right\rangle_{\text{CSVS}} = &\frac{1}{4m^{2 }\mathcal{G}^6(\mathbf{t})} \biggr[3+\Upsilon ^{*4}+\Upsilon ^4+6\Upsilon ^{*2}\Upsilon
^2+12\Upsilon ^*\Upsilon\nonumber\\
&-6\Upsilon ^{*2}-6\Upsilon ^2-4\Upsilon ^{*3}\Upsilon -4\Upsilon ^*\Upsilon ^3+12\text{Sinh}^4\rho
+12\text{Cosh}^2\rho\text{Sinh}^2\rho \nonumber\\
&+24\text{Cosh$\rho
$} \text{Sinh}^3\rho +12\text{Cosh$\rho $} \text{Sinh$\rho $}\left\{1+2\Upsilon ^*\Upsilon\right\}\nonumber\\
&-\Upsilon ^{*2}-\Upsilon
^2+12\text{Sinh}^2\rho\left\{1+2\Upsilon ^*\Upsilon -\Upsilon ^{*2}-\Upsilon ^2\right\}\biggr].
\end{align}

Using Eqs. (\ref{3.5}-\ref{3.6}, \ref{3.11}-\ref{3.19}) values of \(\left\langle :\overset{\wedge }\Phi ^2\overset{\wedge }\Pi ^2:\right\rangle_{\text{CSVS}}\) is

\begin{align} \label{5.5}
\left\langle :\overset{\wedge }\Phi ^2\overset{\wedge }\Pi ^2:\right\rangle _{\text{CSVS}}=&\frac{1}{4m^{2 }\mathbf{t}^2}\biggr[3+\Upsilon ^{*4}+\Upsilon ^4+6\Upsilon ^{*2}\Upsilon
^2+12\Upsilon ^*\Upsilon -6\Upsilon ^{*2}\nonumber\\
&-6\Upsilon ^2-4\Upsilon ^{*3}\Upsilon -4\Upsilon ^*\Upsilon ^3+12\text{Sinh}^4\rho
+12\text{Cosh}^2\rho \text{Sinh}^2\rho\nonumber\\
&+24\text{Cosh$\rho
$} \text{Sinh}^3\rho+12\text{Cosh$\rho $} \text{Sinh$\rho $}\left\{1+2\Upsilon ^*\Upsilon-\Upsilon ^{*2}-\Upsilon
^2\right\} \nonumber\\
&+12\text{Sinh}^2\rho \left\{1+2\Upsilon ^*\Upsilon -\Upsilon ^{*2}-\Upsilon ^2\right\}\biggr].
\end{align}

Using Eqs. (\ref{3.5}-\ref{3.6}, \ref{3.11}-\ref{3.19}) values of \(\left\langle :\overset{\wedge }\Pi ^2\overset{\wedge }\Phi ^2:\right\rangle {}_{\text{CSVS}}\) is 

\begin{align} \label{5.6}
\left\langle :\overset{\wedge }\Pi ^2\overset{\wedge }\Phi ^2:\right\rangle _{\text{CSVS}}=&\frac{1}{4m^{2 }\mathbf{t}^2}\biggr[3+\Upsilon ^{*4}+\Upsilon ^4+6\Upsilon ^{*2}\Upsilon
^2+12\Upsilon ^*\Upsilon\nonumber\\
&-6\Upsilon ^{*2}-6\Upsilon ^2-4\Upsilon ^{*3}\Upsilon -4\Upsilon ^*\Upsilon ^3+12\text{Sinh}^4\rho\nonumber\\
&+12\text{Cosh}^2\rho\text{Sinh}^2\rho +24\text{Cosh$\rho
$}\text{Sinh}^3\rho\nonumber\\
&+12\text{Cosh$\rho $}\text{Sinh$\rho $}2\left\{1+2\Upsilon ^*\Upsilon -\Upsilon ^{*2}-\Upsilon
^2\right\}\nonumber\\
&+12\text{Sinh}^2\rho \left\{1+2\Upsilon ^*\Upsilon -\Upsilon ^{*2}-\Upsilon ^2\right\}\biggr].
\end{align}

Using (\ref{5.3}-\ref{5.6}) in Eq. (\ref{5.2}), the $\left\langle :T_{00}^2:\right\rangle _{\text{CSVS}}$ is 

\begin{align} \label{5.7}
\left\langle :T_{00}^2:\right\rangle _{\text{CSVS}}=&\left(\frac{1}{16m^{2}\mathbf{t}^4}+\frac{1}{8\mathbf{t}^2}+\frac{m^{2}}{16}\right)\biggr[3+\Upsilon
^{*4}+\Upsilon ^4\nonumber\\
&+6\Upsilon ^{*2}\Upsilon ^2+12\Upsilon ^*\Upsilon -6\Upsilon ^{*2}-6\Upsilon ^2-4\Upsilon ^{*3}\Upsilon -4\Upsilon ^*\Upsilon ^3\nonumber\\
&+12\text{Sinh}^4\rho +12\text{Cosh}^2\rho \text{Sinh}^2\rho+24\text{Cosh$\rho $} \text{Sinh}^3\rho \nonumber\\
&+12\text{Cosh$\rho $} \text{Sinh$\rho $}\left\{1+2\Upsilon
^*\Upsilon -\Upsilon ^{*2}-\Upsilon ^2\right\}\nonumber\\
&+12\text{Sinh}^2\rho\left\{1+2\Upsilon ^*\Upsilon -\Upsilon ^{*2}-\Upsilon ^2\right\}\biggr].
\end{align}

Further, $\left\langle :T_{00}:\right\rangle _{\text{CSVS}}$ is given as 

\begin{equation} \label{5.8}
\left\langle :T_{00}:\right\rangle _{\text{CSVS}} =\frac{1}{2}\mathcal{G}^3 (\mathbf{t})m^{2 }\left\langle :\overset{\wedge }\Phi ^2:\right\rangle {}_{\text{CSVS}}+\frac{1}{2\mathcal{G}^3
(\mathbf{t})}\left\langle :\overset{\wedge }\Pi ^2:\right\rangle_{\text{CSVS}}.
\end{equation}

Using Eqs. (\ref{3.5}-\ref{3.6}, \ref{3.11}-\ref{3.19}) values of \(\left\langle :\overset{\wedge }\Pi ^2:\right\rangle {}_{\text{CSVS}}\) is

\begin{align} \label{5.9}
\left\langle :\overset{\wedge }\Pi ^2:\right\rangle {}_{\text{CSVS}}=&\frac{\mathcal{G}^3 (\mathbf{t})}{2m\mathbf{t}^2}[2\text{Sinh}^2\rho
+2\text{Cosh$\rho $} \text{Sinh$\rho $}+1+\Upsilon ^*\Upsilon -\Upsilon ^{*2}-\Upsilon ^2],
\end{align}

Using Eqs. (\ref{3.5}-\ref{3.6}, \ref{3.11}-\ref{3.19}) values of \(\left\langle :\overset{\wedge }\Phi ^2:\right\rangle {}_{\text{CSVS}}\) is 

\begin{align} \label{5.10}
\left\langle :\overset{\wedge }\Phi ^2:\right\rangle_{\text{CSVS}}=&\frac{1}{2\text{m$\mathcal{G}$}^3 (\mathbf{t}}[2\text{Sinh}^2\rho+2\text{Cosh$\rho
$} \text{Sinh$\rho $}+1+\Upsilon ^*\Upsilon -\Upsilon ^{*2}-\Upsilon ^2].
\end{align}

Using Eqs. (\ref{5.9}) and (\ref{5.10}) in Eqs. (\ref{5.8}), the $\left\langle :T_{00}:\right\rangle _{\text{CSVS}}$  is

\begin{align} \label{5.11}
\left\langle :T_{00}:\right\rangle _{\text{CSVS}}=&\left[\frac{m}{4}+\frac{1}{4\text{m$\mathbf{t}$}^2}\right][2\text{Sinh}^2\rho
+2\text{Cosh$\rho $} \text{Sinh$\rho $}+1+\Upsilon ^*\Upsilon -\Upsilon ^{*2}-\Upsilon ^2].
\end{align}

Taking square of (\ref{5.11}) is 

\begin{align} \label{5.12}
\left\langle :T_{00}:\rangle \right.{}^2{}_{\text{CSVS}}=&\left(\frac{1}{16m^{2 }\mathbf{t}^4}+\frac{1}{8\mathbf{t}^2}+\frac{m^{2 }}{16}\right)\biggr[1+\Upsilon
^{*4}+\Upsilon ^4+3\Upsilon ^{*2}\Upsilon ^2\nonumber\\
&+2\Upsilon ^*\Upsilon -2\Upsilon ^{*2}-2\Upsilon ^2-2\Upsilon ^{*3}\Upsilon -2\Upsilon ^*\Upsilon ^3+4\text{Sinh}^4\rho \nonumber\\
&+4\text{Cosh}^2\rho\text{Sinh}^2\rho +8\text{Cosh$\rho $} \text{Sinh}^3\rho \nonumber\\
&+4\text{Cosh$\rho $} \text{Sinh$\rho $}\left\{1+\Upsilon
^*\Upsilon -\Upsilon ^{*2}-\Upsilon ^2\right\}\nonumber\\
&+4\text{Sinh}^2\rho \left\{1+\Upsilon ^*\Upsilon -\Upsilon ^{*2}-\Upsilon ^2\right\}\biggr].
\end{align}

using \(\left\langle :T_{00}^2:\right\rangle {}_{\text{CSVS}}\) and \(\left\langle :T_{00}:\rangle \right.{}^2{}_{\text{CSVS}}\) in Eq. (\ref{5.1}), density fluctuations for CSVS is 

\begin{align} \label{5.13}
\triangle _{\text{CSVS}}=&\left(\frac{1}{16m^{2 }\mathbf{t}^4}+\frac{1}{8\mathbf{t}^2}+\frac{m^{2 }}{16}\right)\biggr[2+3\Upsilon ^{*2}\Upsilon ^2+10\Upsilon
^*\Upsilon -4\Upsilon ^{*2}\nonumber\\
&-4\Upsilon ^2-2\Upsilon ^{*3}\Upsilon -2\Upsilon ^*\Upsilon ^3+8\text{Sinh}^4\rho +8\text{Cosh}^2\rho
 \text{Sinh}^2\rho\nonumber\\
&+16\text{Cosh$\rho $}\text{Sinh}^3\rho+8\text{Cosh$\rho
$}\text{Sinh$\rho $}\left\{1-\varUpsilon ^{*2}-\varUpsilon ^2\right\}\nonumber\\
&+20\text{Cosh$\rho $} \text{Sinh$\rho $}\left\{\varUpsilon ^*\varUpsilon
\right\}+8\text{Sinh}^2\rho\left\{1-\varUpsilon^{*2}-\varUpsilon ^2\right\}\nonumber\\
&+20\text{Sinh}^2\rho \left\{\varUpsilon ^*\varUpsilon \right\}\biggr].   
\end{align}

\begin{figure}[h]%
\centering
\includegraphics[width=0.6\textwidth]{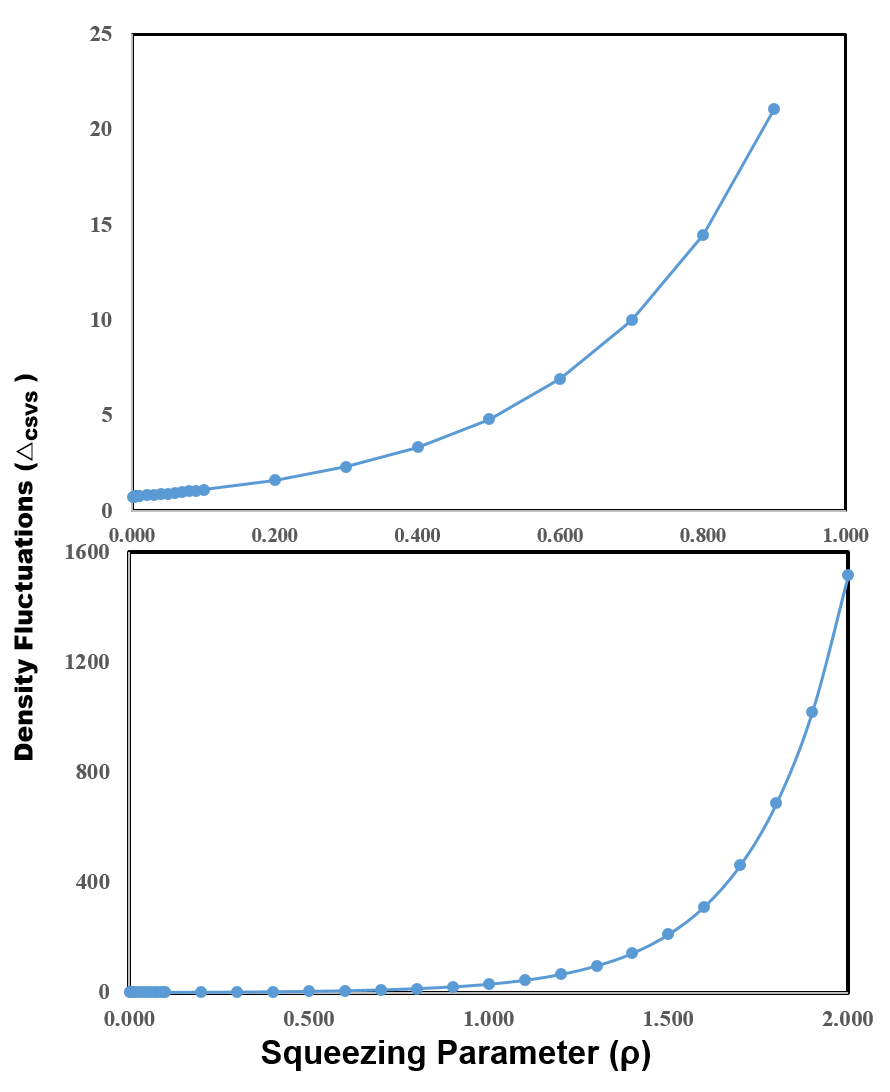}
\caption{ Variation of \(\triangle _{\text{CSVS}}\) with squeezing parameter $\rho $}
\label{fig:figure_1}
\end{figure}

\begin{figure}[h]%
\centering
\includegraphics[width=0.9\textwidth]{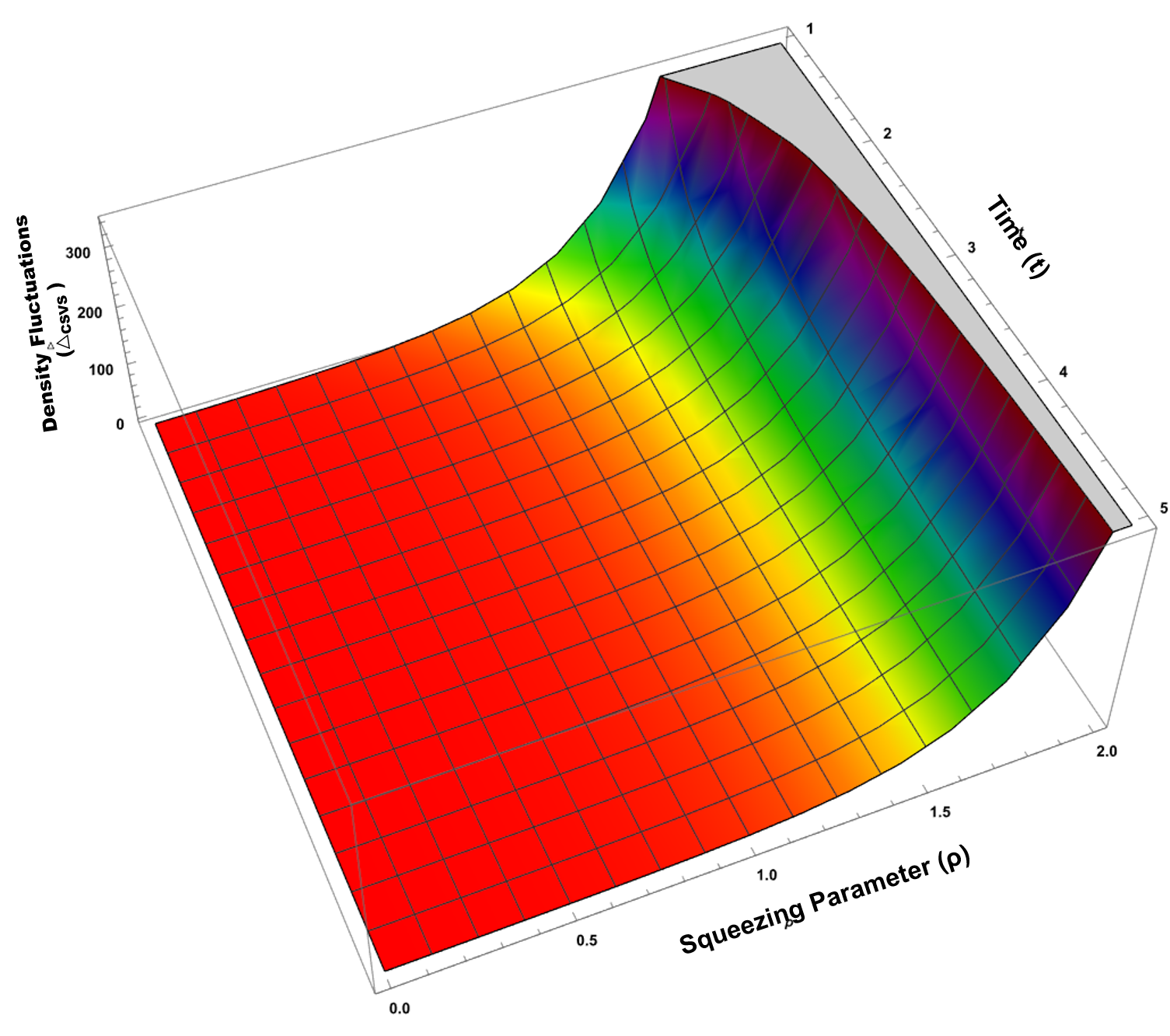}
\caption{3-D plot demonstrating variation in \(\triangle _{\text{CSVS}}\) with squeezing parameter $\rho $ and t.}
\label{fig:figure_2}
\end{figure}

Eq. (\ref{5.13}) showing that \(\triangle _{\text{CSVS}}\) is a function of $\varUpsilon $ and $\rho $, but it strongly depends on Coherent state parameter $\varUpsilon $ than $\rho$. Table \ref{tab:table_1} show \(\triangle _{\text{CSVS}}\) for various squeezing parameter $\rho$ calculated using Eq. (\ref{5.13}). For simplicity of explanation, we use $\varUpsilon=\varUpsilon^*$=1. Similarly for $\varUpsilon^*=\varUpsilon$=0, Eq. (\ref{5.13}) reduces to \(\triangle _{\text{SVS}}\) \cite{venkataratnam_density_2008, venkataratnam_oscillatory_2010, venkataratnam_behavior_2013}, affirming validity of the approximation used in SCEE. Fig.  \ref{fig:figure_1} plotted variation of \(\triangle _{\text{CSVS}}\) with $\rho $, shows increase in \(\triangle _{\text{CSVS}}\) with increasing value of $\rho $. 3-D plot  (Fig.  \ref{fig:figure_2}) showing similar variation of \(\triangle _{\text{CSVS}}\) with $\rho $ and t.

\begin{table}[h]\label{table 1}
\caption{Numerical values of \(\triangle _{\text{CSVS}}\) showing its dependency on
$\rho <<<$1}
\label{tab:table_1}
\begin{tabular}{@{}llllllll@{}}
\toprule
$\rho<<<1$ & \(\triangle _{\text{CSVS}}\)  & $\rho<<1$ & \(\triangle _{\text{CSVS}}\) & $\rho<1$ & \(\triangle _{\text{CSVS}}\) & $\rho$ & \(\triangle _{\text{CSVS}}\)\\
\midrule
0.001 & 0.753005 & 0.010 & 0.780506 & 0.100 & 1.10661 & 1.100 & 44.9879 \\
0.002 & 0.756020 & 0.020 & 0.812049 & 0.200 & 1.60868  & 1.200 & 66.0168\\
0.003 & 0.759045 & 0.030 & 0.844667 & 0.300 & 2.32112 & 1.300 & 97.1180 \\
0.004 & 0.762080 & 0.040 & 0.878399 & 0.400 & 3.33929 & 1.400 & 143.1860 \\
0.005 & 0.765126 & 0.050 & 0.913287 & 0.500 & 4.80367 & 1.500 & 211.5070 \\
0.006 & 0.768181 & 0.060 & 0.949373 & 0.600 & 6.92165 & 1.600 & 312.9390 \\
0.007 & 0.771247 & 0.070 & 0.986702 & 0.700 & 9.99992 & 1.700 & 463.6560 \\
0.008 & 0.774323 & 0.080 & 1.025320 & 0.800 & 14.49280& 1.800 & 687.7640 \\
0.009 & 0.777409 & 0.090 & 1.065270 & 0.900 & 21.07390 & 1.900 & 1021.2000\\
0.010 & 0.780506 & 0.100 & 1.106610 & 1.000 & 30.74360 & 2.000 & 1517.5300\\
\botrule
\end{tabular}
\end{table}

\section{Validity of Uncertainty Relation and Quantum fluctuations in CSVS}
Validity of Uncertainty Relation for CSVS is investigated by determining quantum fluctuations in inflaton field for a coherently oscillating non-classical inflaton. Following dispersion formula of $\Phi$ and $\Pi$ are used to study quantum fluctuations in CSVS as 

\begin{equation} \label{6.1}
 \left\langle :\triangle \hat{\Phi}:\right\rangle_{\text{CSVS}}^2 = \left\langle :\hat{\Phi }^2:\right\rangle_{\text{CSVS}} - \left\langle :\hat{\Phi}:\right\rangle ^2_{\text{CSVS}},
 \end{equation}
 
and

 \begin{equation} \label{6.2}
 \left\langle :\triangle \hat{\Pi}:\right\rangle_{\text{CSVS}}^2 = \left\langle:\hat{\Pi }^2:\right\rangle_{\text{CSVS}} - \left\langle :\hat{\Pi }:\right\rangle ^2{}_{\text{CSVS}},
 \end{equation}

here $\left\langle :\hat{\Phi }^2:\right\rangle$, $\left\langle:\hat{\Pi }^2:\right\rangle$, $\left\langle :\hat{\Phi}:\right\rangle ^2$, and $\left\langle :\hat{\Pi }:\right\rangle ^2$ are the normal ordered expectation values. In order to evaluate dispersion relation for Coherent Squeezed Vacuum State, using  equation (\ref{3.13}) we get $\left\langle :\hat{\Phi}:\right\rangle ^2$ in Coherent Squeezed Vacuum State as

\begin{align} \label{6.3}
\left\langle :\hat{\Phi}:\right\rangle ^2_{\text{CSVS}} =
& \biggr[2\varUpsilon \varUpsilon ^*\Phi \Phi ^*-\varUpsilon ^{*2}\Phi ^{*2}-\varUpsilon
^2\Phi^2\biggr],
\end{align}

using Eqs. (\ref{3.4}-\ref{3.9}) simplifying Eq. (\ref{6.3}) 

\begin{align} \label{6.4}
\left\langle :\hat{\Phi }:\right\rangle ^2{}_{\text{CSVS}} =& \frac{1}{2\text{m$\mathcal{G}$}^3 (\mathbf{t})}\left[2\varUpsilon \varUpsilon ^*-\varUpsilon
^{*2}e^{2\text{im$\mathbf{t}$}}-\varUpsilon ^2e^{-2\text{im$\mathbf{t}$}}\right],
\end{align}

using  equation (\ref{3.16}) we get $\left\langle :\hat{\Pi }:\right\rangle ^2$ in Coherent Squeezed Vacuum State as

\begin{align} \label{6.5}
\left\langle :\hat{\Pi }:\right\rangle ^2{}_{\text{CSVS}} =& \mathcal{G}^6 (\mathbf{t})\left[2\varUpsilon \varUpsilon ^*\dot{\Phi }\dot{\Phi }^*-\varUpsilon
^{*2}\dot{\Phi }^{*2}-\varUpsilon ^2\dot{\Phi }^2\right],
\end{align}

using Eqs. (\ref{3.4}-\ref{3.9}) simplifying Eq. (\ref{6.5}) as

\begin{align} \label{6.6}
\left\langle :\hat{\Pi }:\right\rangle ^2_{\text{CSVS}} = &\frac{\mathcal{G}^3 (\mathbf{t})}{2m}\biggr[2\varUpsilon \varUpsilon ^*\left(m^2+\frac{1}{\mathbf{t}^2}\right)-\varUpsilon ^{*2}e^{2\text{im$\mathbf{t}$}}\left(\frac{1}{ \mathbf{t}^2}-m^2-\frac{2\text{im}}{ \mathbf{t}}\right)\nonumber\\
&-\varUpsilon
^2e^{-2\text{im$\mathbf{t}$}}\left(\frac{1}{ \mathbf{t}^2}-m^2+\frac{2\text{im}}{ \mathbf{t}}\right)\biggr],
\end{align}

using  equation (\ref{3.14}) we get $\left\langle :\hat{\Phi }^2:\right\rangle$ in Coherent Squeezed Vacuum State as

\begin{align} \label{6.7}
\left\langle :\hat{\Phi }^2:\right\rangle_{\text{CSVS}}=& \biggr[\left(2\varUpsilon\varUpsilon ^*+2\text{Sinh}^2\rho +1\right)\Phi\Phi^*-\left(\varUpsilon
^{*2}-e^{-i\Psi }\text{Cosh$\rho $} \text{Sinh$\rho $}\right)\Phi ^{*2}\nonumber\\
&-\left(\varUpsilon ^2-e^{i \Psi }\text{Cosh$\rho $} \text{Sinh$\rho $}\right)\Phi
^2\biggr],
\end{align}

using Eqs. (\ref{3.4}-\ref{3.9}) simplifying Eq. (\ref{6.7}) as

\begin{align} \label{6.8}
\left\langle :\hat{\Phi }^2:\right\rangle {}_{\text{CSVS}} &= \frac{1}{2\text{m$\mathcal{G}$}^3 (\mathbf{t})}\biggr[\left(2 \text{Sinh}^2\rho +1+2\varUpsilon
\varUpsilon ^*\right)-\varUpsilon ^{*2}e^{2\text{im$\mathbf{t}$}}\nonumber\\
&-\varUpsilon ^2e^{-2\text{im$\mathbf{t}$}}+2\text{Cosh$\rho $} \text{Sinh$\rho $cos}(\Psi
-2\text{m$\mathbf{t}$})\biggr],
\end{align}

using equation (\ref{6.4} and \ref{6.8}) in equation (\ref{6.1}) we get $\left\langle :\triangle \hat{\Phi}:\right\rangle_{\text{CSVS}}$

\begin{align} \label{6.9}
\left\langle :\triangle \hat{\Phi}:\right\rangle_{\text{CSVS}} = &\sqrt{\frac{1}{2\text{m$\mathcal{G}$}^3 (\mathbf{t})}\biggr[\left(2 \text{Sinh}^2\rho+1\right)+2\text{Cosh$\rho $} \text{Sinh$\rho $cos}(\Psi -2\text{m$\mathbf{t}$})\biggr]},
\end{align}

using  equation (\ref{3.16}) we get $\left\langle :\hat{\Pi }^2:\right\rangle$ in Coherent Squeezed Vacuum State as

\begin{align} \label{6.10}
\left\langle:\hat{\Pi }^2:\right\rangle_{\text{CSVS}} =& \mathcal{G}^6 (\mathbf{t})\biggr[\left(2\text{Sinh}^2\rho +1+2\varUpsilon \varUpsilon
^*\right)\dot{\Phi }\dot{\Phi }^*\nonumber\\
&-\left(\varUpsilon ^{*2}-e^{-i \Psi }\text{Cosh$\rho$}\text{Sinh$\rho $}\right)\dot{\Phi }^{*2}-\left(\varUpsilon^2-e^{i\Psi }\text{Cosh$\rho $} \text{Sinh$\rho $}\right)\dot{\Phi}^2\biggr],
\end{align}

using Eqs. (\ref{3.4}-\ref{3.9}) simplifying Eq. (\ref{6.10}) as

\begin{align} \label{6.11}
\left\langle:\hat{\Pi }^2:\right\rangle_{\text{CSVS}}=  &\frac{\mathcal{G}^3(\mathbf{t})}{2m}\biggr[\left(2\text{Sinh}^2\rho +1+2\varUpsilon\varUpsilon^*\right)\left(m^2+\frac{1}{ \mathbf{t}^2}\right)\nonumber\\
-&\varUpsilon ^{*2}e^{2\text{im$\mathbf{t}$}}\left(\frac{1}{ \mathbf{t}^2}-m^2-\frac{2\text{im}}{
\mathbf{t}}\right)-\varUpsilon ^2e^{-2\text{im$\mathbf{t}$}}\left(\frac{1}{ \mathbf{t}^2}-m^2+\frac{2\text{im}}{ \mathbf{t}}\right)\nonumber\\
&+2\text{Cosh$\rho
$}\text{Sinh$\rho $}\biggr[\cos (\Psi -2\text{m$\mathbf{t}$})\left(\frac{1}{\mathbf{t}^2}-m^2\right)-\frac{2m}{\mathbf{t}}\sin (\Psi-2\text{m$\mathbf{t}$})\biggr],
\end{align}

using equation (\ref{6.6} and \ref{6.11}) in equation (\ref{6.2}) we get $\left\langle :\triangle \hat{\Pi}:\right\rangle_{\text{CSVS}}$

\begin{align} \label{6.12}
\left\langle :\triangle \hat{\Pi }:\right\rangle _{\text{CSVS}} = &\biggr[\frac{\mathcal{G}^3 (\mathbf{t})}{2m}\biggr[\left(2 \text{Sinh}^2\rho +1\right)\left(m^2+\frac{1}{
\mathbf{t}^2}\right)\nonumber\\
&+2\text{Cosh$\rho $}\text{Sinh$\rho $}\left[\cos (\Psi -2\text{m$\mathbf{t}$})\left(\frac{1}{ \mathbf{t}^2}-m^2\right)-\frac{2m}{\mathbf{t}}\sin (\Psi -2\text{m$\mathbf{t}$})\right]\biggr]\biggr]^\frac{1}{2},
\end{align}

using equation (\ref{6.9} and \ref{6.12}) Quantum fluctuations in inflaton field for Coherent Squeezed Vacuum State given as

\begin{align} \label{6.13}
\left\langle :\triangle \hat{\Phi }:\right\rangle _{\text{CSVS}}\left\langle :\triangle \hat{\Pi }:\right\rangle _{\text{CSVS}} &= \frac{1}{2m}\biggr[\biggr[\left(2
\text{Sinh}^2\rho +1\right)\nonumber\\
&+2\text{Cosh$\rho $} \text{Sinh$\rho $cos}(\Psi -2\text{m$\mathbf{t}$})\biggr]\biggr[\left(2 \text{Sinh}^2\rho +1\right)\left(m^2+\frac{1}{
\mathbf{t}^2}\right)\nonumber\\
&+2\text{Cosh$\rho $} \text{Sinh$\rho $}\biggr(\cos (\Psi -2\text{m$\mathbf{t}$})\left(\frac{1}{ \mathbf{t}^2}-m^2\right)\nonumber\\
&-\frac{2m}{
\mathbf{t}}\sin (\Psi -2\text{m$\mathbf{t}$})\biggr)\biggr]\biggr]^\frac{1}{2}.
\end{align}

\begin{figure}[h]%
\centering
\includegraphics[width=0.9\textwidth]{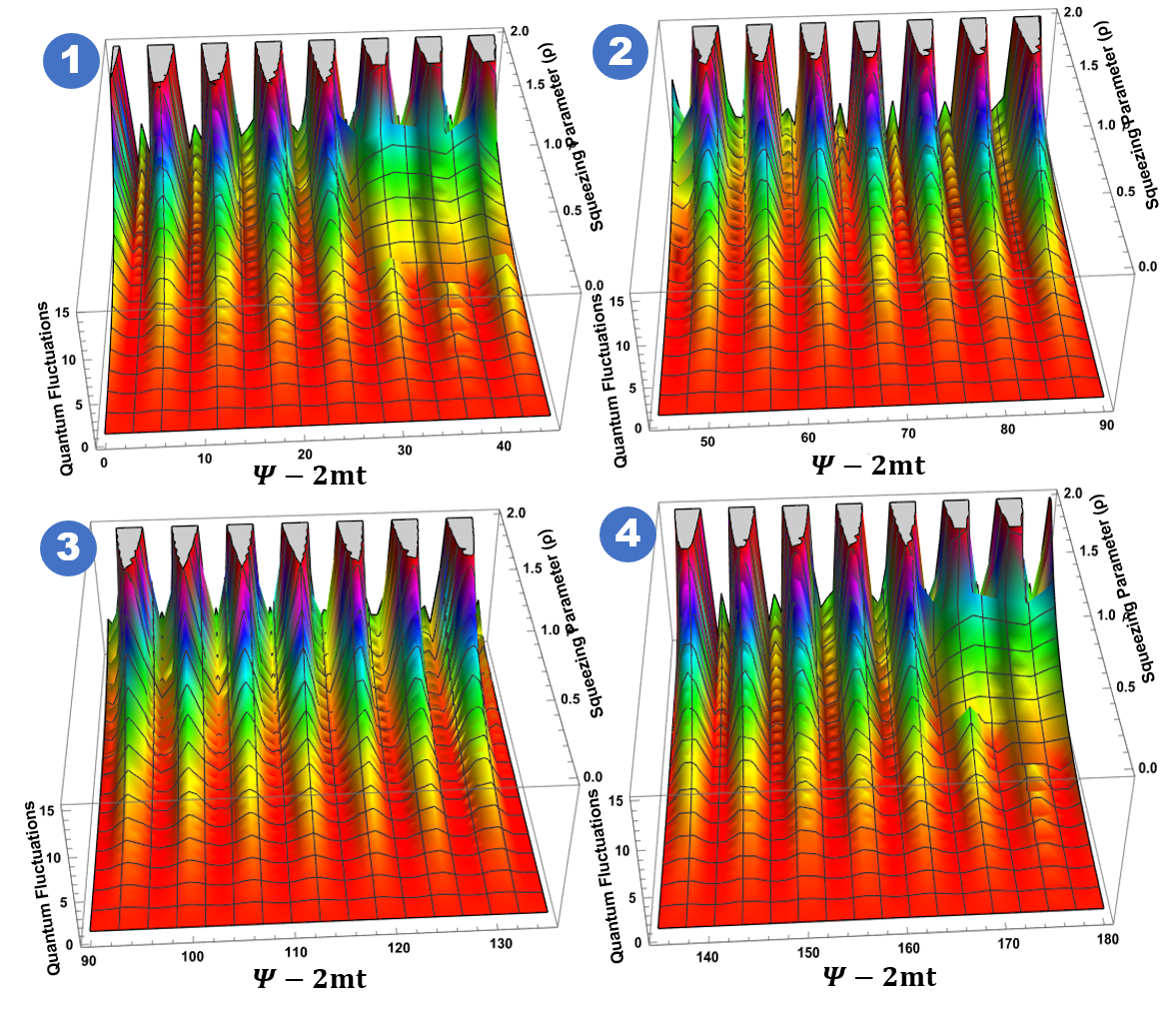}
\caption{3-D plot showing variation of Quantum Fluctuations $\left\langle :\triangle \hat{\Phi }:\right\rangle _{\text{CSVS}}\left\langle :\triangle \hat{\Pi }:\right\rangle _{\text{CSVS}}$ with the values of squeezing parameter $\rho $ and $(\Psi -2m\mathbf{t})$. Where in (1) $(\Psi -2m\mathbf{t})$ is varying between (0$^{\circ}$-45$^{\circ}$), (2) $(\Psi -2m\mathbf{t})$ is varying between (45$^{\circ}$-90$^{\circ}$), (3) $(\Psi -2m\mathbf{t})$ is varying between (90$^{\circ}$-135$^{\circ}$), (4) $(\Psi -2m\mathbf{t})$ is varying between (135$^{\circ}$-180$^{\circ}$).}
\label{fig:figure_3}
\end{figure}

\begin{figure}[h]%
\centering
\includegraphics[width=0.9\textwidth]{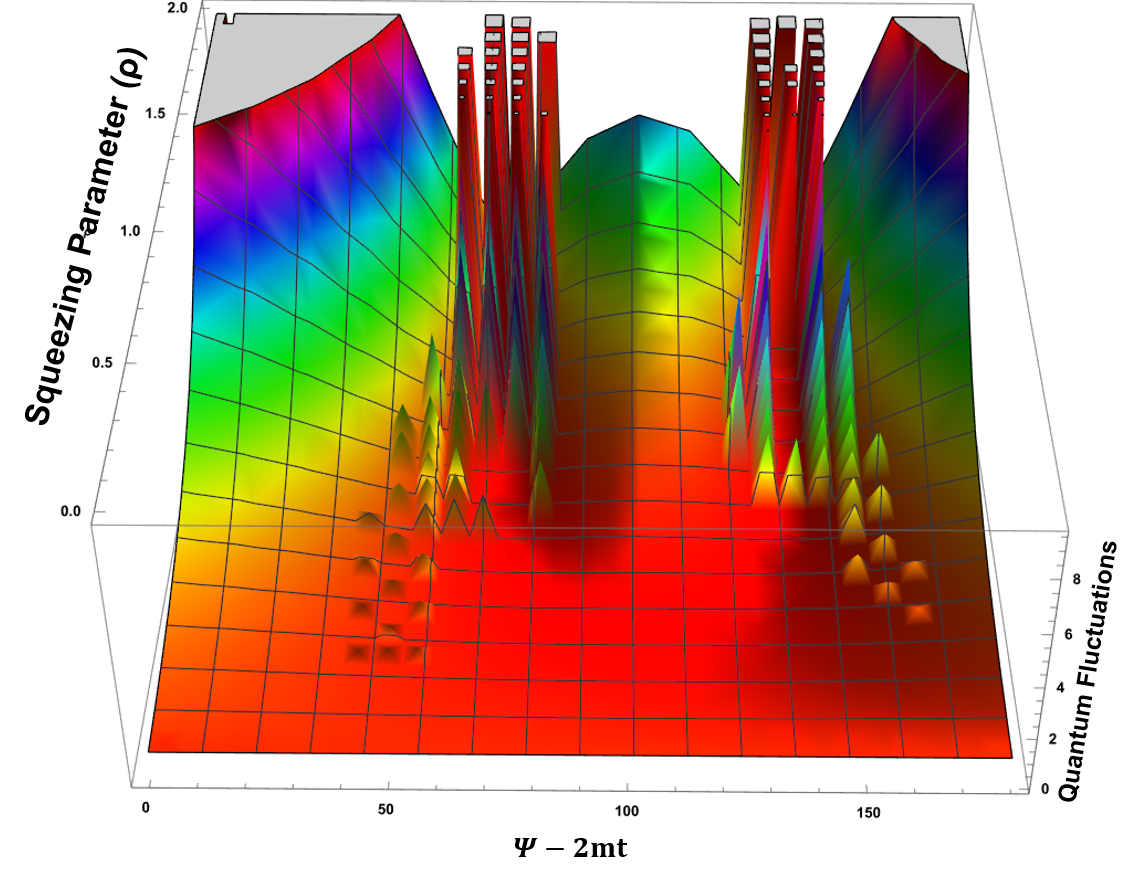}
\caption{A 3-D plot, demonstrating variation of $\left\langle :\triangle \hat{\Phi }:\right\rangle _{\text{CSVS}}\left\langle :\triangle \hat{\Pi }:\right\rangle _{\text{CSVS}}$ with $\rho $ and $(\Psi -2m\mathbf{t})$.}
\label{fig:figure_4}
\end{figure}

Equation (\ref{6.13}) represents the Quantum fluctuations in inflaton field for Coherent Squeezed Vacuum State. Equation (\ref{6.13}) show that, Quantum fluctuations doesn't depend on coherent parameter $\Upsilon$ as uncertainty relation doesn't effected by displacing it with any amount $\Upsilon$ in phase space. As shown in table (\ref{tab:table_2}-\ref{tab:table_5}), calculated values of Quantum fluctuations for all values of $(\Psi -2m\mathbf{t})$  is greater than $\frac{1}{2}$ according to uncertainty relation. The fact can be better understand using 3-D plot (\ref{fig:figure_3}- \ref{fig:figure_4}) between Quantum Fluctuations, $\rho $ and $(\Psi -2m\mathbf{t})$.

\begin{table}[h]\label{table 2}
\caption{Calculated values of quantum fluctuations $\left\langle :\triangle \hat{\Phi }:\right\rangle _{\text{CSVS}}\left\langle :\triangle \hat{\Pi }:\right\rangle _{\text{CSVS}}$ for various values of $\rho$ at $(\Psi -2m\mathbf{t})$=0$^{\circ}$}
\label{tab:table_2}
\begin{tabular}{@{}llllllll@{}}
\toprule
$\rho$ & $(\triangle \hat{\Phi })(\triangle \hat{\Pi })$  & $\rho$ & $(\triangle \hat{\Phi })(\triangle \hat{\Pi })$ & $\rho$ & $(\triangle \hat{\Phi })(\triangle \hat{\Pi })$ & $\rho$ & $(\triangle \hat{\Phi })(\triangle \hat{\Pi })$\\
\midrule
0.001 & 0.707815 & 0.010 & 0.714285 & 0.100 & 0.789276 & 1.100 & 4.540120 \\
0.002 & 0.708525 & 0.020 & 0.721680 & 0.200 & 0.897990 & 1.200 & 5.534220 \\
0.003 & 0.709238 & 0.030 & 0.729297 & 0.300 & 1.039240 & 1.300 & 6.750410 \\
0.004 & 0.709952 & 0.040 & 0.737142 & 0.400 & 1.219940 & 1.400 & 8.237510 \\
0.005 & 0.710669 & 0.050 & 0.745219 & 0.500 & 1.448190 & 1.500 & 10.05520 \\
0.006 & 0.711388 & 0.060 & 0.762091 & 0.600 & 1.733720 & 1.600 & 12.27650 \\
0.007 & 0.712109 & 0.070 & 0.762091 & 0.700 & 2.088340 & 1.700 & 14.99040 \\
0.008 & 0.712832 & 0.080 & 0.770897 & 0.800 & 2.526490 & 1.800 & 18.30590 \\
0.009 & 0.713557 & 0.090 & 0.779957 & 0.900 & 3.065870 & 1.900 & 22.35620 \\
0.010 & 0.714285 & 0.100 & 0.789276 & 1.000 & 3.728210 & 2.000 & 27.30370 \\
\botrule
\end{tabular}
\end{table}

\begin{table}[h]\label{table 3}
\caption{Calculated values of quantum fluctuations $\left\langle :\triangle \hat{\Phi }:\right\rangle _{\text{CSVS}}\left\langle :\triangle \hat{\Pi }:\right\rangle _{\text{CSVS}}$ for various values of $\rho$ at $(\Psi -2m\mathbf{t})$=45$^{\circ}$}
\label{tab:table_3}
\begin{tabular}{@{}llllllll@{}}
\toprule
$\rho$ & $(\triangle \hat{\Phi })(\triangle \hat{\Pi })$  & $\rho$ & $(\triangle \hat{\Phi })(\triangle \hat{\Pi })$ & $\rho$ & $(\triangle \hat{\Phi })(\triangle \hat{\Pi })$ & $\rho$ & $(\triangle \hat{\Phi })(\triangle \hat{\Pi })$\\
\midrule
0.001 & 0.707107 & 0.010 & 0.707177 & 0.100 & 0.714237 & 1.100 & 2.338060 \\
0.002 & 0.707110 & 0.020 & 0.707390 & 0.200 & 0.736329 & 1.200 & 2.823130 \\
0.003 & 0.707113 & 0.030 & 0.707744 & 0.300 & 0.775457 & 1.300 & 3.421270 \\
0.004 & 0.707118 & 0.040 & 0.708240 & 0.400 & 0.834977 & 1.400 & 4.156590 \\
0.005 & 0.707124 & 0.050 & 0.708878 & 0.500 & 0.919392 & 1.500 & 5.058650 \\
0.006 & 0.707132 & 0.060 & 0.709660 & 0.600 & 1.034230 & 1.600 & 6.163700 \\
0.007 & 0.707141 & 0.070 & 0.710586 & 0.700 & 1.186010 & 1.700 & 7.516090 \\
0.008 & 0.707152 & 0.080 & 0.711656 & 0.800 & 1.382340 & 1.800 & 9.170120 \\
0.009 & 0.707164 & 0.090 & 0.712873 & 0.900 & 1.632220 & 1.900 & 11.19220 \\
0.010 & 0.707177 & 0.100 & 0.714237 & 1.000 & 1.946430 & 2.000 & 13.66340 \\
\botrule
\end{tabular}
\end{table}

\begin{table}[h]\label{table 4}
\caption{Calculated values of quantum fluctuations $\left\langle :\triangle \hat{\Phi }:\right\rangle _{\text{CSVS}}\left\langle :\triangle \hat{\Pi }:\right\rangle _{\text{CSVS}}$ for various values of $\rho$ at $(\Psi -2m\mathbf{t})$=60$^{\circ}$}
\label{tab:table_4}
\begin{tabular}{@{}llllllll@{}}
\toprule
$\rho$ & $(\triangle \hat{\Phi })(\triangle \hat{\Pi })$  & $\rho$ & $(\triangle \hat{\Phi })(\triangle \hat{\Pi })$ & $\rho$ & $(\triangle \hat{\Phi })(\triangle \hat{\Pi })$ & $\rho$ & $(\triangle \hat{\Phi })(\triangle \hat{\Pi })$\\
\midrule
0.001 & 0.706849 & 0.010 & 0.704594 & 0.100 & 0.688410 & 1.100 & 1.551170 \\
0.002 & 0.706592 & 0.020 & 0.702229 & 0.200 & 0.683059 & 1.200 & 1.847200 \\
0.003 & 0.706337 & 0.030 & 0.700011 & 0.300 & 0.690503 & 1.300 & 2.216610 \\
0.004 & 0.706084 & 0.040 & 0.697937 & 0.400 & 0.711506 & 1.400 & 2.674540 \\
0.005 & 0.705832 & 0.050 & 0.696004 & 0.500 & 0.748120 & 1.500 & 3.239550 \\
0.006 & 0.705581 & 0.060 & 0.694212 & 0.600 & 0.803609 & 1.600 & 3.934440 \\
0.007 & 0.705332 & 0.070 & 0.692558 & 0.700 & 0.882334 & 1.700 & 4.787150 \\
0.008 & 0.705084 & 0.080 & 0.691041 & 0.800 & 0.989638 & 1.800 & 5.831940 \\
0.009 & 0.704838 & 0.090 & 0.689659 & 0.900 & 1.131800 & 1.900 & 7.110770 \\
0.010 & 0.704594 & 0.100 & 0.688410 & 1.000 & 1.316120 & 2.000 & 8.674960 \\
\botrule
\end{tabular}
\end{table}

\begin{table}[h]\label{table 5}
\caption{Calculated values of quantum fluctuations $\left\langle :\triangle \hat{\Phi }:\right\rangle _{\text{CSVS}}\left\langle :\triangle \hat{\Pi }:\right\rangle _{\text{CSVS}}$ for various values of $\rho$ at $(\Psi -2m\mathbf{t})$=180$^{\circ}$}
\label{tab:table_5}
\begin{tabular}{@{}llllllll@{}}
\toprule
$\rho$ & $(\triangle \hat{\Phi })(\triangle \hat{\Pi })$  & $\rho$ & $(\triangle \hat{\Phi })(\triangle \hat{\Pi })$ & $\rho$ & $(\triangle \hat{\Phi })(\triangle \hat{\Pi })$ & $\rho$ & $(\triangle \hat{\Phi })(\triangle \hat{\Pi })$\\
\midrule
0.001 & 0.706401 & 0.010 & 0.700141 & 0.100 & 0.646204 & 1.100 & 0.503060 \\
0.002 & 0.705697 & 0.020 & 0.693382 & 0.200 & 0.601940 & 1.200 & 0.502053 \\
0.003 & 0.704995 & 0.030 & 0.686826 & 0.300 & 0.570350 & 1.300 & 0.501377 \\
0.004 & 0.704295 & 0.040 & 0.680467 & 0.400 & 0.548155 & 1.400 & 0.500924 \\
0.005 & 0.703598 & 0.050 & 0.674302 & 0.500 & 0.532761 & 1.500 & 0.500619 \\
0.006 & 0.702902 & 0.060 & 0.668324 & 0.600 & 0.522187 & 1.600 & 0.500415 \\
0.007 & 0.702209 & 0.070 & 0.662530 & 0.700 & 0.514978 & 1.700 & 0.500278 \\
0.008 & 0.701517 & 0.080 & 0.656915 & 0.800 & 0.510089 & 1.800 & 0.500187 \\
0.009 & 0.700828 & 0.090 & 0.651475 & 0.900 & 0.506785 & 1.900 & 0.500125 \\
0.010 & 0.700141 & 0.100 & 0.646204 & 1.000 & 0.504558 & 2.000 & 0.500084 \\
\botrule
\end{tabular}
\end{table}

\section{Particle Production of inflaton in CSVS}

For this study, we also investigated production of particle for inflaton in CSVS within context of semiclassical gravity. Here, we calculate No. of particles created at some time t in vacuum state with reference to initial time $t_0$ is 

\begin{align} \label {7.1}
\mathcal{N}_n\left(\mathbf{t},\mathbf{t}_0\right) &= <0,\Phi ,\mathbf{t}_0\left|\hat{\mathcal{N}}(\mathbf{t})\right|0,\Phi ,\mathbf{t}_0>,
\end{align}
 
 here $\hat{\mathcal{N}}(\mathbf{t}) = e^{\dagger}e$. Using Eqs. (\ref{3.11}-\ref{3.12}) the number operator and normal order expectation value of it can be written as

\begin{equation} \label{7.2}
<:\hat{\mathcal{N}}(\mathbf{t}):> = \Phi\Phi^*<:\hat{\Pi}^2:> -\mathcal{G}^3\Phi\dot{\Phi}^*<:\hat{\Pi}\hat{\Phi}:>-\mathcal{G}^3\dot{\Phi}\Phi^*<:\hat{\Phi}\hat{\Pi}:> + \mathcal{G}^6\dot{\Phi}\dot{\Phi}^*<:\hat{\Phi}^2:>. 
\end{equation}

Using Eqs. (\ref{3.5}-\ref{3.6}, \ref{3.11}-\ref{3.19}), the $ \left\langle :\hat{\Pi }\hat{\Phi }:\right\rangle $ for CSVS at initial time $t_0$ is

\begin{align} \label{7.3}
\left\langle :\hat{\Pi }\hat{\Phi }:\right\rangle _{\text{CSVS}} =& \mathcal{G}^3 (\mathbf{t})\biggr[\left(\varUpsilon \varUpsilon ^*+\text{Cosh}^2\rho
\right)\dot{\Phi }\left(\mathbf{t}_0\right)\Phi ^*\left(\mathbf{t}_0\right)\nonumber\\&-\left(\varUpsilon ^2-e^{i \Psi }\text{Cosh$\rho $} \text{Sinh$\rho $}\right)\dot{\Phi}
\left(\mathbf{t}_0\right)\Phi \left(\mathbf{t}_0\right)\nonumber\\&-\left(\varUpsilon ^{*2}-e^{-i \Psi }\text{Cosh$\rho $} \text{Sinh$\rho $}\right)\dot{\Phi}^*\left(\mathbf{t}_0\right)\Phi ^*\left(\mathbf{t}_0\right)\nonumber\\&+\left(\text{Sinh}^2\rho +\varUpsilon \varUpsilon ^*\right)\dot{\Phi }^*\left(\mathbf{t}_0\right)\Phi
\left(\mathbf{t}_0\right)\biggr],
\end{align}

using Eqs. (\ref{3.5}-\ref{3.6}, \ref{3.11}-\ref{3.19}), the $ \left\langle :\hat{\Phi }\hat{\Pi }:\right\rangle $ for CSVS at initial time $t_0$ is

\begin{align} \label{7.4}
\left\langle :\hat{\Phi }\hat{\Pi }:\right\rangle _{\text{CSVS}} =& \mathcal{G}^3 (\mathbf{t})\biggr[\left(\varUpsilon \varUpsilon ^*+\text{Cosh}^2\rho
\right)\Phi \left(\mathbf{t}_0\right)\dot{\Phi }^*\left(\mathbf{t}_0\right)\nonumber\\&-\left(\varUpsilon ^2-e^{i \Psi }\text{Cosh$\rho $} \text{Sinh$\rho $}\right)\dot{\Phi
}\left(\mathbf{t}_0\right)\Phi \left(\mathbf{t}_0\right)\nonumber\\&-\left(\varUpsilon ^{*2}-e^{-i \Psi }\text{Cosh$\rho $} \text{Sinh$\rho $}\right)\dot{\Phi
}^*\left(\mathbf{t}_0\right)\Phi ^*\left(\mathbf{t}_0\right)\nonumber\\&+\left(\text{Sinh}^2\rho +\varUpsilon \varUpsilon ^*\right)\dot{\Phi }\left(\mathbf{t}_0\right)\Phi
^*\left(\mathbf{t}_0\right)\biggr],
\end{align}

using Eqs. (\ref{3.5}-\ref{3.6}, \ref{3.11}-\ref{3.19}), the $ \left\langle :\hat{\Phi }^2:\right\rangle $ for CSVS at initial time $t_0$ is

\begin{align} \label{7.5}
\left\langle :\hat{\Phi }^2:\right\rangle {}_{\text{CSVS}} =& \biggr[\left(2\varUpsilon \varUpsilon ^*+2 \text{Sinh}^2\rho +1\right)\Phi \left(\mathbf{t}_0\right)\Phi
^*\left(\mathbf{t}_0\right)\nonumber\\&-\left(\varUpsilon ^{*2}-e^{-i \Psi }\text{Cosh$\rho $} \text{Sinh$\rho $}\right)\Phi ^*\left(\mathbf{t}_0\right)\Phi
^*\left(\mathbf{t}_0\right)\nonumber\\&-\left(\varUpsilon ^2-e^{i \Psi }\text{Cosh$\rho $} \text{Sinh$\rho $}\right)\Phi \left(\mathbf{t}_0\right)\Phi \left(\mathbf{t}_0\right)\biggr],
\end{align}

using Eqs. (\ref{3.5}-\ref{3.6}, \ref{3.11}-\ref{3.19}), the $ \left\langle :\hat{\Pi }^2:\right\rangle $ for CSVS at initial time $t_0$ is

\begin{align} \label{7.6}
\left\langle :\hat{\Pi }^2:\right\rangle {}_{\text{CSVS}} =& \mathcal{G}^6 (\mathbf{t})\biggr[\left(2 \text{Sinh}^2\rho +1+2\varUpsilon \varUpsilon
^*\right)\dot{\Phi }\left(\mathbf{t}_0\right)\dot{\Phi }^*\left(\mathbf{t}_0\right)\nonumber\\&-\left(\varUpsilon ^{*2}-e^{-i \Psi }\text{Cosh$\rho $} \text{Sinh$\rho
$}\right)\dot{\Phi }^*\left(\mathbf{t}_0\right)\dot{\Phi }^*\left(\mathbf{t}_0\right)\nonumber\\&-\left(\varUpsilon ^2-e^{i \Psi }\text{Cosh$\rho $} \text{Sinh$\rho
$}\right)\dot{\Phi }\dot{\left(\mathbf{t}_0\right)\Phi }\left(\mathbf{t}_0\right)\biggr],
\end{align}

Substituting the values of $ \left\langle :\hat{\Pi }\hat{\Phi }:\right\rangle $, $ \left\langle :\hat{\Phi }\hat{\Pi }:\right\rangle $, $ \left\langle :\hat{\Phi }^2:\right\rangle $  and $ \left\langle :\hat{\Pi }^2:\right\rangle $
in Eq. (\ref{7.2}) then the $\left\langle:\hat{\mathcal{N}}(\mathbf{t}):\right\rangle_{\text{CSVS}}$ is

\begin{align} \label{7.7}
\left\langle:\hat{\mathcal{N}}(\mathbf{t}):\right\rangle_{\text{CSVS}} =& \mathcal{G}^6(\mathbf{t})\left(2 \text{Sinh}^2\rho +1+2\varUpsilon \varUpsilon
^*\right)|\Phi(\mathbf{t})\dot{\Phi}(\mathbf{t}_0)-\dot{\Phi}(\mathbf{t})\Phi(\mathbf{t}_0)|^2\nonumber\\&
 + \left(\text{Sinh}^2\rho +\varUpsilon \varUpsilon
^*\right)
+\mathcal{G}^6(\mathbf{t})\biggr(e^{-i \Psi }\text{Cosh$\rho $} \text{Sinh$\rho
$}\nonumber\\&-\varUpsilon ^{*2}\biggr)\biggr(\Phi(\mathbf{t})\Phi^*(\mathbf{t})\dot{\Phi}^{*}(\mathbf{t}_0)\dot{\Phi}^{*}(\mathbf{t}_0) -\Phi(\mathbf{t})\dot{\Phi}^*(\mathbf{t})\dot{\Phi}^*(\mathbf{t}_0)\Phi^*(\mathbf{t}_0) \nonumber\\&- \dot{\Phi}(\mathbf{t})\Phi^*(\mathbf{t})\Phi^*(\mathbf{t}_0)\dot{\Phi}^*(\mathbf{t}_0) + \dot{\Phi}(\mathbf{t})\dot{\Phi}^*(\mathbf{t})\Phi^{*}(\mathbf{t}_0)\Phi^{*}(\mathbf{t}_0)\biggr) \nonumber\\&
 +\mathcal{G}^6(\mathbf{t})\biggr(e^{i \Psi }\text{Cosh$\rho $} \text{Sinh$\rho
$}-\varUpsilon ^2\biggr)\biggr(\Phi(\mathbf{t})\Phi^*(\mathbf{t})\dot{\Phi}(\mathbf{t}_0)\dot{\Phi}(\mathbf{t}_0) \nonumber\\&-\Phi(\mathbf{t})\dot{\Phi}^*(\mathbf{t})\dot{\Phi}(\mathbf{t}_0)\Phi(\mathbf{t}_0) - \dot{\Phi}(\mathbf{t})\Phi^*(\mathbf{t})\Phi(\mathbf{t}_0)\dot{\Phi}(\mathbf{t}_0) \nonumber\\&+ \dot{\Phi}(\mathbf{t})\dot{\Phi}^*(\mathbf{t})\Phi(\mathbf{t}_0)\Phi(\mathbf{t}_0)\biggr).
\end{align}

Here for simplicity, considering in equation (\ref{7.7}) as
\begin{equation} \label{7.8}
\mathcal{N}_0(\mathbf{t},\mathbf{t}_0) = \mathcal{G}^6(\mathbf{t})|\Phi(\mathbf{t})\dot{\Phi}(\mathbf{t}_0)-\dot{\Phi}(\mathbf{t})\Phi(\mathbf{t}_0)|^2,
\end{equation}

\begin{align} \label{7.10}
\mathbf{Q} =& \biggr(\Phi(\mathbf{t})\Phi^*(\mathbf{t})\dot{\Phi}^{*}(\mathbf{t}_0)\dot{\Phi}^{*}(\mathbf{t}_0) -\Phi(\mathbf{t})\dot{\Phi}^*(\mathbf{t})\dot{\Phi}^*(\mathbf{t}_0)\Phi^*(\mathbf{t}_0) \nonumber\\&- \dot{\Phi}(\mathbf{t})\Phi^*(\mathbf{t})\Phi^*(\mathbf{t}_0)\dot{\Phi}^*(\mathbf{t}_0) + \dot{\Phi}(\mathbf{t})\dot{\Phi}^*(\mathbf{t})\Phi^{*}(\mathbf{t}_0)\Phi^{*}(\mathbf{t}_0)\biggr),
\end{align}

\begin{align} \label{7.11}
\mathbf\mathcal{R} =& \biggr(\Phi(\mathbf{t})\Phi^*(\mathbf{t})\dot{\Phi}(\mathbf{t}_0)\dot{\Phi}(\mathbf{t}_0) -\Phi(\mathbf{t})\dot{\Phi}^*(\mathbf{t})\dot{\Phi}(\mathbf{t}_0)\Phi(\mathbf{t}_0) \nonumber\\&- \dot{\Phi}(\mathbf{t})\Phi^*(\mathbf{t})\Phi(\mathbf{t}_0)\dot{\Phi}(\mathbf{t}_0) + \dot{\Phi}(\mathbf{t})\dot{\Phi}^*(\mathbf{t})\Phi(\mathbf{t}_0)\Phi(\mathbf{t}_0)\biggr),
\end{align}

Eqs. (\ref{7.8}-\ref{7.11}) are substituted in Eq.(\ref{7.7}) then we get

\begin{align} \label{7.12}
\left\langle:\hat{\mathcal{N}}(\mathbf{t}):\right\rangle_{\text{CSVS}} =& (2 \text{Sinh}^2\rho +1+2\varUpsilon \varUpsilon
^*)\mathcal{N}_0(\mathbf{t},\mathbf{t}_0) + (\text{Sinh}^2\rho +\varUpsilon \varUpsilon
^*)\nonumber\\&
+\mathcal{G}^6(\mathbf{t})(e^{-i \Psi }\text{Cosh$\rho $} \text{Sinh$\rho
$}-\varUpsilon ^{*2})\mathbf{Q} +\mathcal{G}^6(\mathbf{t})(e^{i \Psi }\text{Cosh$\rho $} \text{Sinh$\rho
$}-\varUpsilon ^2)\mathbf\mathcal{R}.
\end{align}

Using  Eqs. (\ref{4.15}-\ref{4.18}) in Eqs. (\ref{7.10}-\ref{7.11}), $\mathbf{Q}$ and $\mathbf{R}$ can be calculated as

\begin{align} \label{7.13}
\mathbf{Q} &= \frac{1}{4\chi(\mathbf{t})\chi(t_0)\mathcal{G}^3(\mathbf{t})\mathcal{G}^3(\mathbf{t_0})}\biggr[exp(2i\int\chi(t_0)dt_0)\biggr(\frac{9}{4}\biggr(\frac{\dot{\mathcal{G}}(\mathbf{t_0})}{\mathcal{G}(\mathbf{t_0})}\biggr)^2 + \frac{1}{4}\biggr(\frac{\dot{\chi}(\mathbf{t_0})}{\chi(\mathbf{t_0})}\biggr)^2 \nonumber\\& +\frac{3}{2}\frac{\dot{\mathcal{G}}(\mathbf{t_0})}{\mathcal{G}(\mathbf{t_0})}\frac{\dot{\chi}(\mathbf{t_0})}{\chi(\mathbf{t_0})} -\chi^2(t_0) - i\chi(t_0)\biggr(\frac{3\dot{\mathcal{G}}(\mathbf{t_0})}{\mathcal{G}(\mathbf{t_0})} +\frac{\dot{\chi}(t_0)}{\chi(t_0)} \biggr)+ \frac{9}{4}\biggr(\frac{\dot{\mathcal{G}}(\mathbf{t})}{\mathcal{G}(\mathbf{t})}\biggr)^2 \nonumber\\& + \frac{1}{4}\biggr(\frac{\dot{\chi}(\mathbf{t})}{\chi(\mathbf{t})}\biggr)^2+\frac{3}{2}\frac{\dot{\mathcal{G}}(\mathbf{t})}{\mathcal{G}(\mathbf{t})}\frac{\dot{\chi}(\mathbf{t})}{\chi(\mathbf{t})}+\chi^2(\mathbf{t})\biggr) \nonumber\\&
-2\biggr(\frac{3}{4}\frac{\dot{\mathcal{G}}(\mathbf{t_0})}{\mathcal{G}(\mathbf{t_0})} +\frac{\dot{\chi}(t_0)}{4\chi(t_0)} -\frac{i}{2}\chi(t_0) \biggr)\biggr(\frac{3\dot{\mathcal{G}}(\mathbf{t})}{\mathcal{G}(\mathbf{t})} + \frac{\dot{\chi}(\mathbf{t})}{\chi(\mathbf{t})}\biggr) \biggr], 
\end{align}

\begin{align} \label{7.14}
\mathbf\mathcal{R} &= \frac{exp(-2i\int\chi(t_0)dt_0)}{4\chi(\mathbf{t})\chi(t_0)\mathcal{G}^3(\mathbf{t})\mathcal{G}^3(\mathbf{t_0})}\biggr[\biggr(\frac{9}{4}\biggr(\frac{\dot{\mathcal{G}}(\mathbf{t_0})}{\mathcal{G}(\mathbf{t_0})}\biggr)^2 + \frac{1}{4}\biggr(\frac{\dot{\chi}(\mathbf{t_0})}{\chi(\mathbf{t_0})}\biggr)^2\nonumber\\&+\frac{3}{2}\frac{\dot{\mathcal{G}}(\mathbf{t_0})}{\mathcal{G}(\mathbf{t_0})}\frac{\dot{\chi}(\mathbf{t_0})}{\chi(\mathbf{t_0})} -\chi^2(t_0) + i\chi(t_0)\biggr(\frac{3\dot{\mathcal{G}}(\mathbf{t_0})}{\mathcal{G}(\mathbf{t_0})} +\frac{\dot{\chi}(t_0)}{\chi(t_0)} \biggr)\nonumber\\&+ \frac{9}{4}\biggr(\frac{\dot{\mathcal{G}}(\mathbf{t})}{\mathcal{G}(\mathbf{t})}\biggr)^2 + \frac{1}{4}\biggr(\frac{\dot{\chi}(\mathbf{t})}{\chi(\mathbf{t})}\biggr)^2+\frac{3}{2}\frac{\dot{\mathcal{G}}(\mathbf{t})}{\mathcal{G}(\mathbf{t})}\frac{\dot{\chi}(\mathbf{t})}{\chi(\mathbf{t})}+\chi^2(\mathbf{t})\biggr) \nonumber\\&
-\biggr(\frac{3}{4}\frac{\dot{\mathcal{G}}(\mathbf{t})}{\mathcal{G}(\mathbf{t})} +\frac{\dot{\chi}(\mathbf{t})}{4\chi(\mathbf{t})} -\frac{i}{2}\chi(\mathbf{t}) \biggr)\biggr(\frac{3\dot{\mathcal{G}}(\mathbf{t_0})}{\mathcal{G}(\mathbf{t_0})} + \frac{\dot{\chi}(t_0)}{\chi(t_0)}+2i\chi(t_0)\biggr)\nonumber\\&
-\biggr(\frac{3}{4}\frac{\dot{\mathcal{G}}(\mathbf{t_0})}{\mathcal{G}(\mathbf{t_0})} +\frac{\dot{\chi}(t_0)}{4\chi(t_0)} +\frac{i}{2}\chi(t_0) \biggr)\biggr(\frac{3\dot{\mathcal{G}}(\mathbf{t})}{\mathcal{G}(\mathbf{t})} + \frac{\dot{\chi}(\mathbf{t})}{\chi(\mathbf{t})}+2i\chi(\mathbf{t}))\biggr) \biggr],
\end{align}

using approximation ansatz (\ref{4.21}-\ref{4.22}) in Eqs.(\ref{7.13}-\ref{7.14}), we get
 \begin{align} \label{7.15}
\mathbf{Q} & ={\exp(2i\int md\mathbf{t}_0)\over 4m^2 \mathcal{G}_o^6\mathbf{t}^2 \mathbf{t}^2_0}\left[{1\over
\mathbf{t}^2_0}+{1\over \mathbf{t}^2}-{2\over \mathbf{t} \mathbf{t}_{0}}-{2im\over \mathbf{t}_0}+{2im\over \mathbf{t}}\right], 
\end{align}

\begin{align} \label{7.16}
\mathbf\mathcal{R} 
&={\exp(-2i\int md\mathbf{t}_0)\over 4m^2\mathcal{G}_0^6\mathbf{t}^2\mathbf{t}^2_0}\left[{1\over
\mathbf{t}^2_0} - {2\over \mathbf{t}\mathbf{t}_0}+{1\over \mathbf{t}^2}+{2im\over \mathbf{t}_0}-{2im\over \mathbf{t}}\right].
\end{align}

Use $\Upsilon= |\Upsilon|e^{i\theta}$, $\varphi=2mt_0$, $\theta=mt_0$  with Eqs. (\ref{7.15}) and (\ref{7.16}) in Eq. (\ref{7.12}) we get

\begin{align} \label{7.17}
\left\langle:\hat{\mathcal{N}}(\mathbf{t}):\right\rangle_{\text{CSVS}} &\simeq (2\sinh^2 \rho +\Upsilon ^*\Upsilon +1)\frac{(\mathbf{t}-\mathbf{t}_0)^2}{4m^2\mathbf{t}_0^4}\nonumber\\& 
 + (\sinh^2\rho +\Upsilon ^*\Upsilon ) 
 +\biggr(\sinh2\rho-\frac{\Upsilon ^*\Upsilon}{2}\biggr)\frac{(\mathbf{t}-\mathbf{t}_0)^2}{4m^2\mathbf{t}_0^4}. 
\end{align}

The Eq. (\ref{7.17}) show that number of particles produced in CSVS.

\section{Results and Discussion}

In this paper, we have conducted an analysis of the quantum effects on the Friedmann-Robertson-Walker universe, focusing on Power-law Expansion, Density Fluctuations, Quantum Fluctuations, and Production of Particles of inflaton for CSVS within framework of semiclassical theory of gravity. We emphasize the significant role of the non-classical state of gravity, particularly in particle production, while considering background metric as classical with quantize matter field \cite{anderson_effects_1983, campos_semiclassical_1994}. \\
Initially, we have determined oscillatory phase of inflaton for CSVS. We have analyzed that density fluctuations for CSVS depends on $\rho$ as well as $\varUpsilon$ that can be reduced to squeezed vacuum state \cite{venkataratnam_density_2008} for $\varUpsilon=0$, demonstrate sustainability of SCTG and inflaton energy density \cite{lachieze-rey_cosmological_1999,takahashi_thermo_1996, xu_quantum_2007}. The $\triangle _{\text{CSVS}}$ exhibit an inverse proportionality to various powers of time t as demonstrated by the SCEE model. Fig. \ref{fig:figure_1} illustrates density fluctuations for CSVS, depicts an increase in density fluctuations as the squeezing parameter $\rho$ increase.\\
In our analysis, we adopted a minimal coupling approach between gravity and inflaton to examine importance of massive inflaton within context of an isotropic and homogeneous universe. Additionally, we derived the leading approximate solution to SCEE for scale factor in CSVS. Notably, scale factor is found to be a function of the squeezing parameter $\rho$, coherent angle $\theta$, angle of squeezing $\varphi$ and coherent state parameter $\Upsilon$. When we set limit $m\mathbf{t} >> 1$, $\rho = 0$, $\theta = m\mathbf{t}$ and $\varphi = 2m\mathbf{t}$, the scale factor  $\mathcal{G}_1(\mathbf{t})_{CSVS}$ varies as $t^{2/3}$, i.e., the study revealed that SCEE of gravity provides similar power law expansion of scale factor as obtained by classical theory.\\
We have also studied quantum fluctuation for CSVS. Our findings in the matter indicates that the dispersion of field is inversely proportional to time, while momentum dispersion is proportional to time. Using Eq. (\ref{6.13}), different quantum states produce values according to uncertainty relation, show validity of approximation used. Hence, CSVS holds significant physical importance in preserving the quantum properties of the coherently oscillating inflaton field. This preservation ensures that the oscillating inflaton field retains quantum characteristics.\\
Additionally, we have investigated the particle production in the CSVS of an oscillating massive inflaton within the framework of semiclassical gravity, specifically within the context of the isotropic and homogeneous universe. We computed the particle production for CSVS as shown in Eq. (\ref{7.17}) that depends on $\Upsilon$, $\rho$, $\varphi$ and $\theta$. Production of particles in CSVS depends on their respective parameters, this implies that these parameters have a considerable influence during the early universe period. Therefore, the states under consideration play a crucial role in the study of the early universe.\\

\bibliography{sn-bibliography.bib}

\begin{thebibliography}{10}
\expandafter\ifx\csname url\endcsname\relax
  \def\url#1{\burl{#1}}\fi
\expandafter\ifx\csname urlprefix\endcsname\relax\def\urlprefix{URL }\fi
\providecommand{\bibinfo}[2]{#2}
\providecommand{\eprint}[2][]{\url{#2}}
\providecommand{\doi}[1]{\url{https://doi.org/#1}}
\bibcommenthead

\bibitem{kubik_origin_2022}
\bibinfo{author}{Kubik, B.}, \bibinfo{author}{Karska, A.} \& \bibinfo{author}{Opitom, C.}
\newblock \bibinfo{title}{Origin of the universe and planetary systems}  (\bibinfo{year}{2022}).
\newblock \urlprefix\url{https://books.rsc.org/books/edited-volume/2003/chapter/4583607/Origin-of-the-Universe-and-Planetary-Systems}.

\bibitem{moore_big_2014}
\bibinfo{author}{Moore, B.} \& \bibinfo{author}{Moore, B.}
\newblock \bibinfo{title}{The big bang}.
\newblock \emph{\bibinfo{journal}{Elephants in Space: The Past, Present and Future of Life and the Universe}} \bibinfo{pages}{43--55} (\bibinfo{year}{2014}).
\newblock \urlprefix\url{https://link.springer.com/chapter/10.1007/978-3-319-05672-2_3}.

\bibitem{guth1981cosmological}
\bibinfo{author}{Guth, A.~H.} \& \bibinfo{author}{Weinberg, E.~J.}
\newblock \bibinfo{title}{Cosmological consequences of a first-order phase transition in the s u 5 grand unified model}.
\newblock \emph{\bibinfo{journal}{Physical Review D}} \textbf{\bibinfo{volume}{23}}, \bibinfo{pages}{876} (\bibinfo{year}{1981}).

\bibitem{green_cosmological_2022}
\bibinfo{author}{Green, D.}, \bibinfo{author}{Guo, Y.} \& \bibinfo{author}{Wallisch, B.}
\newblock \bibinfo{title}{Cosmological implications of axion-matter couplings}.
\newblock \emph{\bibinfo{journal}{Journal of Cosmology and Astroparticle Physics}} \textbf{\bibinfo{volume}{2022}}, \bibinfo{pages}{019} (\bibinfo{year}{2022}).
\newblock \urlprefix\url{https://iopscience.iop.org/article/10.1088/1475-7516/2022/02/019/meta}.

\bibitem{albrecht_inflation_1994}
\bibinfo{author}{Albrecht, A.}, \bibinfo{author}{Ferreira, P.}, \bibinfo{author}{Joyce, M.} \& \bibinfo{author}{Prokopec, T.}
\newblock \bibinfo{title}{Inflation and squeezed quantum states}.
\newblock \emph{\bibinfo{journal}{Physical Review D}} \textbf{\bibinfo{volume}{50}}, \bibinfo{pages}{4807--4820} (\bibinfo{year}{1994}).
\newblock \urlprefix\url{https://link.aps.org/doi/10.1103/PhysRevD.50.4807}.

\bibitem{albrecht_reheating_1982}
\bibinfo{author}{Albrecht, A.}, \bibinfo{author}{Steinhardt, P.~J.}, \bibinfo{author}{Turner, M.~S.} \& \bibinfo{author}{Wilczek, F.}
\newblock \bibinfo{title}{Reheating an {Inflationary} {Universe}}.
\newblock \emph{\bibinfo{journal}{Physical Review Letters}} \textbf{\bibinfo{volume}{48}}, \bibinfo{pages}{1437--1440} (\bibinfo{year}{1982}).

\bibitem{kofman_reheating_1994}
\bibinfo{author}{Kofman, L.}, \bibinfo{author}{Linde, A.} \& \bibinfo{author}{Starobinsky, A.~A.}
\newblock \bibinfo{title}{Reheating after {Inflation}}.
\newblock \emph{\bibinfo{journal}{Physical Review Letters}} \textbf{\bibinfo{volume}{73}}, \bibinfo{pages}{3195--3198} (\bibinfo{year}{1994}).

\bibitem{allahverdi_reheating_2010}
\bibinfo{author}{Allahverdi, R.}, \bibinfo{author}{Brandenberger, R.}, \bibinfo{author}{Cyr-Racine, F.-Y.} \& \bibinfo{author}{Mazumdar, A.}
\newblock \bibinfo{title}{Reheating in {Inflationary} {Cosmology}: {Theory} and {Applications}}.
\newblock \emph{\bibinfo{journal}{Annual Review of Nuclear and Particle Science}} \textbf{\bibinfo{volume}{60}}, \bibinfo{pages}{27--51} (\bibinfo{year}{2010}).

\bibitem{cook_reheating_2015}
\bibinfo{author}{Cook, J.~L.}, \bibinfo{author}{Dimastrogiovanni, E.}, \bibinfo{author}{Easson, D.~A.} \& \bibinfo{author}{Krauss, L.~M.}
\newblock \bibinfo{title}{Reheating predictions in single field inflation}.
\newblock \emph{\bibinfo{journal}{Journal of Cosmology and Astroparticle Physics}} \textbf{\bibinfo{volume}{2015}}, \bibinfo{pages}{047} (\bibinfo{year}{2015}).
\newblock \urlprefix\url{https://dx.doi.org/10.1088/1475-7516/2015/04/047}.

\bibitem{yadav_reheating_2023}
\bibinfo{author}{Yadav, S.}, \bibinfo{author}{Goswami, R.}, \bibinfo{author}{Venkataratnam, K.~K.} \& \bibinfo{author}{Yajnik, U.~A.}
\newblock \bibinfo{title}{Reheating constraints on modified quadratic chaotic inflation}.
\newblock \urlprefix\url{http://arxiv.org/abs/2309.06990}.
\newblock \eprint{2309.06990 [astro-ph]}.

\bibitem{yadav2024reheating}
\bibinfo{author}{Yadav, S.}, \bibinfo{author}{Gangal, D.} \& \bibinfo{author}{Venkataratnam, K.~K.}
\newblock \bibinfo{title}{Reheating constraints on mutated hilltop inflation} (\bibinfo{year}{2024}).
\newblock \eprint{2401.09806}.

\bibitem{mohajan_friedmann_2013}
\bibinfo{author}{Mohajan, H.}
\newblock \bibinfo{title}{Friedmann, {Robertson}-{Walker} ({FRW}) {Models} in {Cosmology}} (\bibinfo{year}{2013}).
\newblock \urlprefix\url{https://mpra.ub.uni-muenchen.de/52402/}.

\bibitem{suresh_particle_2004}
\bibinfo{author}{Suresh, P.~K.}
\newblock \bibinfo{title}{Particle {Creation} in the {Oscillatory} {Phase} of {Inflaton}}.
\newblock \emph{\bibinfo{journal}{International Journal of Theoretical Physics}} \textbf{\bibinfo{volume}{43}}, \bibinfo{pages}{425--436} (\bibinfo{year}{2004}).
\newblock \urlprefix\url{http://link.springer.com/10.1023/B:IJTP.0000028875.07382.4e}.

\bibitem{kim_one-parameter_1999}
\bibinfo{author}{Kim, J.~K.} \& \bibinfo{author}{Kim, S.~P.}
\newblock \bibinfo{title}{One-parameter squeezed {Gaussian} states of a time-dependent harmonic oscillator and the selection rule for vacuum states}.
\newblock \emph{\bibinfo{journal}{Journal of Physics A: Mathematical and General}} \textbf{\bibinfo{volume}{32}}, \bibinfo{pages}{2711--2718} (\bibinfo{year}{1999}).
\newblock \urlprefix\url{https://iopscience.iop.org/article/10.1088/0305-4470/32/14/012}.

\bibitem{finelli_quantum_1999}
\bibinfo{author}{Finelli, F.}, \bibinfo{author}{Gruppuso, A.} \& \bibinfo{author}{Venturi, G.}
\newblock \bibinfo{title}{Quantum fields in an expanding universe}.
\newblock \emph{\bibinfo{journal}{Classical and Quantum Gravity}} \textbf{\bibinfo{volume}{16}}, \bibinfo{pages}{3923--3935} (\bibinfo{year}{1999}).
\newblock \urlprefix\url{https://iopscience.iop.org/article/10.1088/0264-9381/16/12/310}.

\bibitem{geralico_novel_2004}
\bibinfo{author}{Geralico, A.}, \bibinfo{author}{Landolfi, G.}, \bibinfo{author}{Ruggeri, G.} \& \bibinfo{author}{Soliani, G.}
\newblock \bibinfo{title}{Novel approach to the study of quantum effects in the early {Universe}}.
\newblock \emph{\bibinfo{journal}{Physical Review D}} \textbf{\bibinfo{volume}{69}}, \bibinfo{pages}{043504} (\bibinfo{year}{2004}).
\newblock \urlprefix\url{https://link.aps.org/doi/10.1103/PhysRevD.69.043504}.

\bibitem{padmanabhan_gravity_2005}
\bibinfo{author}{Padmanabhan, T.}
\newblock \bibinfo{title}{Gravity and the thermodynamics of horizons}.
\newblock \emph{\bibinfo{journal}{Physics Reports}} \textbf{\bibinfo{volume}{406}}, \bibinfo{pages}{49--125} (\bibinfo{year}{2005}).
\newblock \urlprefix\url{https://linkinghub.elsevier.com/retrieve/pii/S0370157304004582}.

\bibitem{mahajan_particle_2008}
\bibinfo{author}{Mahajan, G.} \& \bibinfo{author}{Padmanabhan, T.}
\newblock \bibinfo{title}{Particle creation, classicality and related issues in quantum field theory: {I}. {Formalism} and toy models}.
\newblock \emph{\bibinfo{journal}{General Relativity and Gravitation}} \textbf{\bibinfo{volume}{40}}, \bibinfo{pages}{661--708} (\bibinfo{year}{2008}).
\newblock \urlprefix\url{http://link.springer.com/10.1007/s10714-007-0526-z}.

\bibitem{lachieze-rey_cosmic_1995}
\bibinfo{author}{Lachièze-Rey, M.} \& \bibinfo{author}{Luminet, J.-P.}
\newblock \bibinfo{title}{Cosmic topology}.
\newblock \emph{\bibinfo{journal}{Physics Reports}} \textbf{\bibinfo{volume}{254}}, \bibinfo{pages}{135--214} (\bibinfo{year}{1995}).
\newblock \urlprefix\url{https://linkinghub.elsevier.com/retrieve/pii/037015739400085H}.

\bibitem{ellis_cosmological_1998}
\bibinfo{author}{Ellis, G. F.~R.} \& \bibinfo{author}{van Elst, H.}
\newblock \bibinfo{title}{Cosmological models ({Cargèse} lectures 1998)}  (\bibinfo{year}{1998}).
\newblock \urlprefix\url{https://arxiv.org/abs/gr-qc/9812046}.

\bibitem{carvalho_scalar_2004}
\bibinfo{author}{Carvalho, A. D.~M.}, \bibinfo{author}{Furtado, C.} \& \bibinfo{author}{Pedrosa, I.~A.}
\newblock \bibinfo{title}{Scalar fields and exact invariants in a {Friedmann}-{Robertson}-{Walker} spacetime}.
\newblock \emph{\bibinfo{journal}{Physical Review D}} \textbf{\bibinfo{volume}{70}}, \bibinfo{pages}{123523} (\bibinfo{year}{2004}).
\newblock \urlprefix\url{https://link.aps.org/doi/10.1103/PhysRevD.70.123523}.

\bibitem{bak1998quantum}
\bibinfo{author}{Bak, D.}, \bibinfo{author}{Kim, S.~P.}, \bibinfo{author}{Kim, S.~K.}, \bibinfo{author}{Soh, K.-S.} \& \bibinfo{author}{Yee, J.~H.}
\newblock \bibinfo{title}{Quantum inflaton dynamics}.
\newblock \emph{\bibinfo{journal}{Physical Review D}} \textbf{\bibinfo{volume}{59}}, \bibinfo{pages}{027301} (\bibinfo{year}{1998}).

\bibitem{kim1999thermal}
\bibinfo{author}{Kim, S.} \& \bibinfo{author}{Page, D.}
\newblock \bibinfo{title}{Thermal inflation model in semiclassical quantum gravity.}
\newblock \emph{\bibinfo{journal}{Journal of Korean Physical Society}} \textbf{\bibinfo{volume}{35}}, \bibinfo{pages}{S660--S665} (\bibinfo{year}{1999}).

\bibitem{guth1985quantum}
\bibinfo{author}{Guth, A.~H.} \& \bibinfo{author}{Pi, S.-Y.}
\newblock \bibinfo{title}{Quantum mechanics of the scalar field in the new inflationary universe}.
\newblock \emph{\bibinfo{journal}{Physical Review D}} \textbf{\bibinfo{volume}{32}}, \bibinfo{pages}{1899} (\bibinfo{year}{1985}).

\bibitem{habib1992stochastic}
\bibinfo{author}{Habib, S.}
\newblock \bibinfo{title}{Stochastic inflation: Quantum phase-space approach}.
\newblock \emph{\bibinfo{journal}{Physical Review D}} \textbf{\bibinfo{volume}{46}}, \bibinfo{pages}{2408} (\bibinfo{year}{1992}).

\bibitem{linde1994big}
\bibinfo{author}{Linde, A.}, \bibinfo{author}{Linde, D.} \& \bibinfo{author}{Mezhlumian, A.}
\newblock \bibinfo{title}{From the big bang theory to the theory of a stationary universe}.
\newblock \emph{\bibinfo{journal}{Physical Review D}} \textbf{\bibinfo{volume}{49}}, \bibinfo{pages}{1783} (\bibinfo{year}{1994}).

\bibitem{kennard_zur_1927}
\bibinfo{author}{Kennard, E.~H.}
\newblock \bibinfo{title}{Zur {Quantenmechanik} einfacher {Bewegungstypen}}.
\newblock \emph{\bibinfo{journal}{Zeitschrift for Physik}} \textbf{\bibinfo{volume}{44}}, \bibinfo{pages}{326--352} (\bibinfo{year}{1927}).
\newblock \urlprefix\url{http://link.springer.com/10.1007/BF01391200}.

\bibitem{venkataratnam_particle_2004}
\bibinfo{author}{Venkataratnam, K.~K.} \& \bibinfo{author}{Suresh, P.~K.}
\newblock \bibinfo{title}{{PARTICLE} {PRODUCTION} {OF} {COHERENTLY} {OSCILLATING} {NONCLASSICAL} {INFLATON} {IN} {FRW} {UNIVERSE}}.
\newblock \emph{\bibinfo{journal}{International Journal of Modern Physics D}} \textbf{\bibinfo{volume}{13}}, \bibinfo{pages}{239--252} (\bibinfo{year}{2004}).
\newblock \urlprefix\url{https://www.worldscientific.com/doi/abs/10.1142/S0218271804004578}.

\bibitem{bakke_geometric_2009}
\bibinfo{author}{Bakke, K.}, \bibinfo{author}{Pedrosa, I.~A.} \& \bibinfo{author}{Furtado, C.}
\newblock \bibinfo{title}{Geometric phases and squeezed quantum states of relic gravitons}.
\newblock \emph{\bibinfo{journal}{Journal of Mathematical Physics}} \textbf{\bibinfo{volume}{50}}, \bibinfo{pages}{113521} (\bibinfo{year}{2009}).
\newblock \urlprefix\url{https://pubs.aip.org/aip/jmp/article/930980}.

\bibitem{pedrosa_gaussian_2009}
\bibinfo{author}{Pedrosa, I.}, \bibinfo{author}{Bakke, K.} \& \bibinfo{author}{Furtado, C.}
\newblock \bibinfo{title}{Gaussian wave packet states of relic gravitons}.
\newblock \emph{\bibinfo{journal}{Physics Letters B}} \textbf{\bibinfo{volume}{671}}, \bibinfo{pages}{314--317} (\bibinfo{year}{2009}).
\newblock \urlprefix\url{https://linkinghub.elsevier.com/retrieve/pii/S0370269308014822}.

\bibitem{stoica_friedmann-lemaitre-robertson-walker_2016}
\bibinfo{author}{Stoica, O.~C.}
\newblock \bibinfo{title}{The {Friedmann}-{Lemaître}-{Robertson}-{Walker} {Big} {Bang} {Singularities} are {Well} {Behaved}}.
\newblock \emph{\bibinfo{journal}{International Journal of Theoretical Physics}} \textbf{\bibinfo{volume}{55}}, \bibinfo{pages}{71--80} (\bibinfo{year}{2016}).
\newblock \urlprefix\url{http://link.springer.com/10.1007/s10773-015-2634-y}.

\bibitem{hu_anisotropy_1978}
\bibinfo{author}{Hu, B.~L.} \& \bibinfo{author}{Parker, L.}
\newblock \bibinfo{title}{Anisotropy damping through quantum effects in the early universe}.
\newblock \emph{\bibinfo{journal}{Physical Review D}} \textbf{\bibinfo{volume}{17}}, \bibinfo{pages}{933--945} (\bibinfo{year}{1978}).
\newblock \urlprefix\url{https://link.aps.org/doi/10.1103/PhysRevD.17.933}.

\bibitem{dhayal_quantum_2020}
\bibinfo{author}{Dhayal, R.}, \bibinfo{author}{Rathore, M.} \& \bibinfo{author}{Venkataratnam, K.~K.}
\newblock \bibinfo{title}{Quantum fluctuations and particle production in the oscillatory phase of a thermal inflaton in a {FRW} universe}.
\newblock \emph{\bibinfo{journal}{Modern Physics Letters A}} \textbf{\bibinfo{volume}{35}}, \bibinfo{pages}{2050022} (\bibinfo{year}{2020}).
\newblock \urlprefix\url{https://www.worldscientific.com/doi/abs/10.1142/S0217732320500224}.

\bibitem{fischetti_quantum_1979}
\bibinfo{author}{Fischetti, M.~V.}, \bibinfo{author}{Hartle, J.~B.} \& \bibinfo{author}{Hu, B.~L.}
\newblock \bibinfo{title}{Quantum effects in the early universe. {I}. {Influence} of trace anomalies on homogeneous, isotropic, classical geometries}.
\newblock \emph{\bibinfo{journal}{Physical Review D}} \textbf{\bibinfo{volume}{20}}, \bibinfo{pages}{1757--1771} (\bibinfo{year}{1979}).
\newblock \urlprefix\url{https://link.aps.org/doi/10.1103/PhysRevD.20.1757}.

\bibitem{hartle_quantum_1979}
\bibinfo{author}{Hartle, J.~B.} \& \bibinfo{author}{Hu, B.~L.}
\newblock \bibinfo{title}{Quantum effects in the early universe. {II}. {Effective} action for scalar fields in homogeneous cosmologies with small anisotropy}.
\newblock \emph{\bibinfo{journal}{Physical Review D}} \textbf{\bibinfo{volume}{20}}, \bibinfo{pages}{1772--1782} (\bibinfo{year}{1979}).
\newblock \urlprefix\url{https://link.aps.org/doi/10.1103/PhysRevD.20.1772}.

\bibitem{hartle_quantum_1980}
\bibinfo{author}{Hartle, J.~B.} \& \bibinfo{author}{Hu, B.~L.}
\newblock \bibinfo{title}{Quantum effects in the early universe. {III}. {Dissipation} of anisotropy by scalar particle production}.
\newblock \emph{\bibinfo{journal}{Physical Review D}} \textbf{\bibinfo{volume}{21}}, \bibinfo{pages}{2756--2769} (\bibinfo{year}{1980}).
\newblock \urlprefix\url{https://link.aps.org/doi/10.1103/PhysRevD.21.2756}.

\bibitem{hartle_quantum_1981}
\bibinfo{author}{Hartle, J.~B.}
\newblock \bibinfo{title}{Quantum effects in the early universe. {V}. {Finite} particle production without trace anomalies}.
\newblock \emph{\bibinfo{journal}{Physical Review D}} \textbf{\bibinfo{volume}{23}}, \bibinfo{pages}{2121--2128} (\bibinfo{year}{1981}).
\newblock \urlprefix\url{https://link.aps.org/doi/10.1103/PhysRevD.23.2121}.

\bibitem{anderson_effects_1983}
\bibinfo{author}{Anderson, P.}
\newblock \bibinfo{title}{Effects of quantum fields on singularities and particle horizons in the early universe}.
\newblock \emph{\bibinfo{journal}{Physical Review D}} \textbf{\bibinfo{volume}{28}}, \bibinfo{pages}{271--285} (\bibinfo{year}{1983}).
\newblock \urlprefix\url{https://link.aps.org/doi/10.1103/PhysRevD.28.271}.

\bibitem{campos_semiclassical_1994}
\bibinfo{author}{Campos, A.} \& \bibinfo{author}{Verdaguer, E.}
\newblock \bibinfo{title}{Semiclassical equations for weakly inhomogeneous cosmologies}.
\newblock \emph{\bibinfo{journal}{Physical Review D}} \textbf{\bibinfo{volume}{49}}, \bibinfo{pages}{1861--1880} (\bibinfo{year}{1994}).
\newblock \urlprefix\url{https://link.aps.org/doi/10.1103/PhysRevD.49.1861}.

\bibitem{geralico_novel_2004-1}
\bibinfo{author}{Geralico, A.}, \bibinfo{author}{Landolfi, G.}, \bibinfo{author}{Ruggeri, G.} \& \bibinfo{author}{Soliani, G.}
\newblock \bibinfo{title}{Novel approach to the study of quantum effects in the early {Universe}}.
\newblock \emph{\bibinfo{journal}{Physical Review D}} \textbf{\bibinfo{volume}{69}}, \bibinfo{pages}{043504} (\bibinfo{year}{2004}).
\newblock \urlprefix\url{https://link.aps.org/doi/10.1103/PhysRevD.69.043504}.

\bibitem{pedrosa_exact_2007}
\bibinfo{author}{Pedrosa, I.}, \bibinfo{author}{Furtado, C.} \& \bibinfo{author}{Rosas, A.}
\newblock \bibinfo{title}{Exact linear invariants and quantum effects in the early universe}.
\newblock \emph{\bibinfo{journal}{Physics Letters B}} \textbf{\bibinfo{volume}{651}}, \bibinfo{pages}{384--387} (\bibinfo{year}{2007}).
\newblock \urlprefix\url{https://linkinghub.elsevier.com/retrieve/pii/S0370269307007599}.

\bibitem{lopes_gaussian_2009}
\bibinfo{author}{Lopes, C. E.~F.}, \bibinfo{author}{Pedrosa, I.~A.}, \bibinfo{author}{Furtado, C.} \& \bibinfo{author}{De~M.~Carvalho, A.~M.}
\newblock \bibinfo{title}{Gaussian wave packet states of scalar fields in a universe of de {Sitter}}.
\newblock \emph{\bibinfo{journal}{Journal of Mathematical Physics}} \textbf{\bibinfo{volume}{50}}, \bibinfo{pages}{083511} (\bibinfo{year}{2009}).
\newblock \urlprefix\url{https://pubs.aip.org/aip/jmp/article/920134}.

\bibitem{gangal2024density}
\bibinfo{author}{Gangal, D.}, \bibinfo{author}{Yadav, S.} \& \bibinfo{author}{Venkataratnam, K.}
\newblock \bibinfo{title}{Density fluctuations for squeezed number state and coherent squeezed number state in flat frw universe}.
\newblock \emph{\bibinfo{journal}{arXiv preprint arXiv:2402.00432}}  (\bibinfo{year}{2024}).

\bibitem{robertson1936kinematics}
\bibinfo{author}{Robertson, H.~P.}
\newblock \bibinfo{title}{Kinematics and world-structure iii.}
\newblock \emph{\bibinfo{journal}{The Astrophysical Journal}} \textbf{\bibinfo{volume}{83}}, \bibinfo{pages}{257} (\bibinfo{year}{1936}).

\bibitem{shaviv2011did}
\bibinfo{author}{Shaviv, G.}
\newblock \bibinfo{title}{Did edwin hubble plagiarize?}
\newblock \emph{\bibinfo{journal}{arXiv preprint arXiv:1107.0442}}  (\bibinfo{year}{2011}).

\bibitem{zel1971creation}
\bibinfo{author}{Zel'Dovich, Y.~B.} \& \bibinfo{author}{Starobinskij, A.}
\newblock \bibinfo{title}{Creation of particles and vacuum polarization in an anisotropic gravitational field.}
\newblock \emph{\bibinfo{journal}{Zhurnal Eksperimentalnoi i Teoreticheskoi Fiziki}} \textbf{\bibinfo{volume}{61}}, \bibinfo{pages}{2161--2175} (\bibinfo{year}{1971}).

\bibitem{bergstrom2006cosmology}
\bibinfo{author}{Bergstr{\"o}m, L.} \& \bibinfo{author}{Goobar, A.}
\newblock \emph{\bibinfo{title}{Cosmology and particle astrophysics}}  (\bibinfo{publisher}{Springer Science \& Business Media}, \bibinfo{year}{2006}).

\bibitem{ellis1999cosmological}
\bibinfo{author}{Ellis, G.~F.} \& \bibinfo{author}{Van~Elst, H.}
\newblock \bibinfo{title}{Cosmological models: Cargese lectures 1998}.
\newblock \emph{\bibinfo{journal}{Theoretical and Observational Cosmology}} \bibinfo{pages}{1--116} (\bibinfo{year}{1999}).

\bibitem{kuo_semiclassical_1993}
\bibinfo{author}{Kuo, C.-I.} \& \bibinfo{author}{Ford, L.~H.}
\newblock \bibinfo{title}{Semiclassical gravity theory and quantum fluctuations}.
\newblock \emph{\bibinfo{journal}{Physical Review D}} \textbf{\bibinfo{volume}{47}}, \bibinfo{pages}{4510--4519} (\bibinfo{year}{1993}).
\newblock \urlprefix\url{https://link.aps.org/doi/10.1103/PhysRevD.47.4510}.

\bibitem{caves_quantum-mechanical_1981}
\bibinfo{author}{Caves, C.~M.}
\newblock \bibinfo{title}{Quantum-mechanical noise in an interferometer}.
\newblock \emph{\bibinfo{journal}{Physical Review D}} \textbf{\bibinfo{volume}{23}}, \bibinfo{pages}{1693--1708} (\bibinfo{year}{1981}).
\newblock \urlprefix\url{https://link.aps.org/doi/10.1103/PhysRevD.23.1693}.

\bibitem{matacz_coherent_1994}
\bibinfo{author}{Matacz, A.~L.}
\newblock \bibinfo{title}{Coherent state representation of quantum fluctuations in the early {Universe}}.
\newblock \emph{\bibinfo{journal}{Physical Review D}} \textbf{\bibinfo{volume}{49}}, \bibinfo{pages}{788--798} (\bibinfo{year}{1994}).
\newblock \urlprefix\url{https://link.aps.org/doi/10.1103/PhysRevD.49.788}.

\bibitem{suresh_thermal_2001}
\bibinfo{author}{Suresh, P.~K.}
\newblock \bibinfo{title}{{THERMAL} {SQUEEZING} {AND} {DENSITY} {FLUCTUATIONS} {IN} {SEMICLASSICAL} {THEORY} {OF} {GRAVITY}}.
\newblock \emph{\bibinfo{journal}{Modern Physics Letters A}} \textbf{\bibinfo{volume}{16}}, \bibinfo{pages}{707--717} (\bibinfo{year}{2001}).
\newblock \urlprefix\url{https://www.worldscientific.com/doi/abs/10.1142/S0217732301003802}.

\bibitem{suresh_squeezed_1998}
\bibinfo{author}{Suresh, P.~K.} \& \bibinfo{author}{Kuriakose, V.~C.}
\newblock \bibinfo{title}{{SQUEEZED} {STATE} {REPRESENTATION} {OF} {QUANTUM} {FLUCTUATIONS} {AND} {SEMICLASSICAL} {THEORY}}.
\newblock \emph{\bibinfo{journal}{Modern Physics Letters A}} \textbf{\bibinfo{volume}{13}}, \bibinfo{pages}{165--172} (\bibinfo{year}{1998}).
\newblock \urlprefix\url{https://www.worldscientific.com/doi/abs/10.1142/S0217732398000218}.

\bibitem{suresh_nonclassical_2001}
\bibinfo{author}{Suresh, P.~K.}
\newblock \bibinfo{title}{{NONCLASSICAL} {STATE} {REPRESENTATION} {OF} {INFLATON} {AND} {POWER}-{LAW} {EXPANSION} {IN} {FRW} {UNIVERSE}}.
\newblock \emph{\bibinfo{journal}{Modern Physics Letters A}} \textbf{\bibinfo{volume}{16}}, \bibinfo{pages}{2431--2438} (\bibinfo{year}{2001}).
\newblock \urlprefix\url{https://www.worldscientific.com/doi/abs/10.1142/S0217732301005874}.

\bibitem{venkataratnam_nonclassical_2010}
\bibinfo{author}{Venkataratnam, K.~K.} \& \bibinfo{author}{Suresh, P.~K.}
\newblock \bibinfo{title}{{NONCLASSICAL} {SCALAR} {FIELD} {IN} {THE} {FRW} {UNIVERSE}}.
\newblock \emph{\bibinfo{journal}{International Journal of Modern Physics D}} \textbf{\bibinfo{volume}{19}}, \bibinfo{pages}{37--61} (\bibinfo{year}{2010}).
\newblock \urlprefix\url{https://www.worldscientific.com/doi/abs/10.1142/S021827181001621X}.

\bibitem{venkataratnam_density_2008}
\bibinfo{author}{Venkataratnam, K.~K.} \& \bibinfo{author}{Suresh, P.~K.}
\newblock \bibinfo{title}{{DENSITY} {FLUCTUATIONS} {IN} {THE} {OSCILLATORY} {PHASE} {OF} {A} {NONCLASSICAL} {INFLATON} {IN} {THE} {FRW} {UNIVERSE}}.
\newblock \emph{\bibinfo{journal}{International Journal of Modern Physics D}} \textbf{\bibinfo{volume}{17}}, \bibinfo{pages}{1991--2005} (\bibinfo{year}{2008}).
\newblock \urlprefix\url{https://www.worldscientific.com/doi/abs/10.1142/S0218271808013662}.

\bibitem{venkataratnam_oscillatory_2010}
\bibinfo{author}{Venkataratnam, K.~K.} \& \bibinfo{author}{Suresh, P.~K.}
\newblock \bibinfo{title}{{OSCILLATORY} {PHASE} {OF} {NONCLASSICAL} {THERMAL} {INFLATON} {IN} {FRW} {UNIVERSE}}.
\newblock \emph{\bibinfo{journal}{International Journal of Modern Physics D}} \textbf{\bibinfo{volume}{19}}, \bibinfo{pages}{1147--1195} (\bibinfo{year}{2010}).
\newblock \urlprefix\url{https://www.worldscientific.com/doi/abs/10.1142/S0218271810017184}.

\bibitem{venkataratnam_behavior_2013}
\bibinfo{author}{Venkataratnam, K.~K.}
\newblock \bibinfo{title}{{BEHAVIOR} {OF} {NON}-{CLASSICAL} {INFLATON} {IN} {THE} {FRW} {UNIVERSE}}.
\newblock \emph{\bibinfo{journal}{Modern Physics Letters A}} \textbf{\bibinfo{volume}{28}}, \bibinfo{pages}{1350168} (\bibinfo{year}{2013}).
\newblock \urlprefix\url{https://www.worldscientific.com/doi/abs/10.1142/S021773231350168X}.

\bibitem{dhayal_quantum_2020-1}
\bibinfo{author}{Dhayal, R.}, \bibinfo{author}{Rathore, M.} \& \bibinfo{author}{Venkataratnam, K.~K.}
\newblock \bibinfo{title}{Quantum fluctuations and particle production in the oscillatory phase of a thermal inflaton in a {FRW} universe}.
\newblock \emph{\bibinfo{journal}{Modern Physics Letters A}} \textbf{\bibinfo{volume}{35}}, \bibinfo{pages}{2050022} (\bibinfo{year}{2020}).
\newblock \urlprefix\url{https://www.worldscientific.com/doi/abs/10.1142/S0217732320500224}.

\bibitem{lachieze-rey_cosmological_1999}
\bibinfo{author}{Ellis, G. F.~R.} \& \bibinfo{author}{Elst, H.}
\newblock \bibinfo{title}{ in \textit{Cosmological {Models}}} (ed.\bibinfo{editor}{Lachièze-Rey, M.}) \emph{\bibinfo{booktitle}{Theoretical and {Observational} {Cosmology}}} \bibinfo{pages}{1--116} (\bibinfo{publisher}{Springer Netherlands}, \bibinfo{address}{Dordrecht}, \bibinfo{year}{1999}).
\newblock \urlprefix\url{http://link.springer.com/10.1007/978-94-011-4455-1_1}.

\bibitem{takahashi_thermo_1996}
\bibinfo{author}{Takahashi, Y.} \& \bibinfo{author}{Umezawa, H.}
\newblock \bibinfo{title}{{THERMO} {FIELD} {DYNAMICS}}.
\newblock \emph{\bibinfo{journal}{International Journal of Modern Physics B}} \textbf{\bibinfo{volume}{10}}, \bibinfo{pages}{1755--1805} (\bibinfo{year}{1996}).
\newblock \urlprefix\url{https://www.worldscientific.com/doi/abs/10.1142/S0217979296000817}.

\bibitem{xu_quantum_2007}
\bibinfo{author}{Xu, X.-L.}, \bibinfo{author}{Li, H.-Q.} \& \bibinfo{author}{Wang, J.-S.}
\newblock \bibinfo{title}{Quantum fluctuations of mesoscopic {RLC} circuit involving complicated coupling in thermal squeezed state}.
\newblock \emph{\bibinfo{journal}{Physica B: Condensed Matter}} \textbf{\bibinfo{volume}{396}}, \bibinfo{pages}{199--206} (\bibinfo{year}{2007}).
\newblock \urlprefix\url{https://linkinghub.elsevier.com/retrieve/pii/S0921452607002256}.

\bibitem{koh_gravitational_2004}
\bibinfo{author}{Koh, S.}, \bibinfo{author}{Kim, S.~P.} \& \bibinfo{author}{Song, D.~J.}
\newblock \bibinfo{title}{Gravitational {Wave} {Spectrum} in {Inflation} with {Nonclassical} {States}}.
\newblock \emph{\bibinfo{journal}{Journal of High Energy Physics}} \textbf{\bibinfo{volume}{2004}}, \bibinfo{pages}{060--060} (\bibinfo{year}{2004}).
\newblock \urlprefix\url{http://stacks.iop.org/1126-6708/2004/i=12/a=060?key=crossref.cd1ab511cad13ad8cdd715eeafc03f08}.

\bibitem{lopes_gaussian_2009-1}
\bibinfo{author}{Lopes, C. E.~F.}, \bibinfo{author}{Pedrosa, I.~A.}, \bibinfo{author}{Furtado, C.} \& \bibinfo{author}{De~M.~Carvalho, A.~M.}
\newblock \bibinfo{title}{Gaussian wave packet states of scalar fields in a universe of de {Sitter}}.
\newblock \emph{\bibinfo{journal}{Journal of Mathematical Physics}} \textbf{\bibinfo{volume}{50}}, \bibinfo{pages}{083511} (\bibinfo{year}{2009}).
\newblock \urlprefix\url{https://pubs.aip.org/aip/jmp/article/920134}.

\bibitem{lachieze-rey_theoretical_1999}
\bibinfo{editor}{Lachièze-Rey, M.} (ed.) \emph{\bibinfo{title}{Theoretical and observational cosmology: proceedings of the {NATO} {Advanced} {Study} {Institute} on {Theoretical} and {Observational} {Cosmology}, {Cargèse}, {France}, {August} 17 - 29, 1998}} No. \bibinfo{number}{v. 541} in \bibinfo{series}{{NATO} science series {Series} {C}, {Mathematical} and physical sciences} (\bibinfo{publisher}{Kluwer Acad. Publ}, \bibinfo{address}{Dordrecht}, \bibinfo{year}{1999}).

\bibitem{lopes_gaussian_2009-2}
\bibinfo{author}{Lopes, C. E.~F.}, \bibinfo{author}{Pedrosa, I.~A.}, \bibinfo{author}{Furtado, C.} \& \bibinfo{author}{De~M.~Carvalho, A.~M.}
\newblock \bibinfo{title}{Gaussian wave packet states of scalar fields in a universe of de {Sitter}}.
\newblock \emph{\bibinfo{journal}{Journal of Mathematical Physics}} \textbf{\bibinfo{volume}{50}}, \bibinfo{pages}{083511} (\bibinfo{year}{2009}).
\newblock \urlprefix\url{https://pubs.aip.org/aip/jmp/article/920134}.

\bibitem{sinha_[no_2003}
\bibinfo{author}{Sinha, S.}, \bibinfo{author}{Raval, A.} \& \bibinfo{author}{Hu, B.~L.}
\newblock \bibinfo{title}{[{No} title found]}.
\newblock \emph{\bibinfo{journal}{Foundations of Physics}} \textbf{\bibinfo{volume}{33}}, \bibinfo{pages}{37--64} (\bibinfo{year}{2003}).
\newblock \urlprefix\url{http://link.springer.com/10.1023/A:1022815724856}.

\bibitem{shaviv_did_2011}
\bibinfo{author}{Shaviv, G.}
\newblock \bibinfo{title}{Did {Edwin} {Hubble} plagiarize?}  (\bibinfo{year}{2011}).
\newblock \urlprefix\url{https://arxiv.org/abs/1107.0442}.

\bibitem{berger1978classical}
\bibinfo{author}{Berger, B.~K.}
\newblock \bibinfo{title}{Classical analog of cosmological particle creation}.
\newblock \emph{\bibinfo{journal}{Physical Review D}} \textbf{\bibinfo{volume}{18}}, \bibinfo{pages}{4367} (\bibinfo{year}{1978}).

\bibitem{berger1981scalar}
\bibinfo{author}{Berger, B.~K.}
\newblock \bibinfo{title}{Scalar field in the early universe: Coherent-state representation and thermal density matrix}.
\newblock \emph{\bibinfo{journal}{Physical Review D}} \textbf{\bibinfo{volume}{23}}, \bibinfo{pages}{1250} (\bibinfo{year}{1981}).

\bibitem{grishchuk1990squeezed}
\bibinfo{author}{Grishchuk, L.} \& \bibinfo{author}{Sidorov, Y.~V.}
\newblock \bibinfo{title}{Squeezed quantum states of relic gravitons and primordial density fluctuations}.
\newblock \emph{\bibinfo{journal}{Physical Review D}} \textbf{\bibinfo{volume}{42}}, \bibinfo{pages}{3413} (\bibinfo{year}{1990}).

\bibitem{brandenberger1992entropy}
\bibinfo{author}{Brandenberger, R.}, \bibinfo{author}{Mukhanov, V.} \& \bibinfo{author}{Prokopec, T.}
\newblock \bibinfo{title}{Entropy of a classical stochastic field and cosmological perturbations}.
\newblock \emph{\bibinfo{journal}{Physical Review Letters}} \textbf{\bibinfo{volume}{69}}, \bibinfo{pages}{3606} (\bibinfo{year}{1992}).

\bibitem{brandenberger1993entropy}
\bibinfo{author}{Brandenberger, R.}, \bibinfo{author}{Mukhanov, V.} \& \bibinfo{author}{Prokopec, T.}
\newblock \bibinfo{title}{Entropy of the gravitational field}.
\newblock \emph{\bibinfo{journal}{Physical Review D}} \textbf{\bibinfo{volume}{48}}, \bibinfo{pages}{2443} (\bibinfo{year}{1993}).

\bibitem{kuo1993semiclassical}
\bibinfo{author}{Kuo, C.-I.} \& \bibinfo{author}{Ford, L.}
\newblock \bibinfo{title}{Semiclassical gravity theory and quantum fluctuations}.
\newblock \emph{\bibinfo{journal}{Physical Review D}} \textbf{\bibinfo{volume}{47}}, \bibinfo{pages}{4510} (\bibinfo{year}{1993}).

\bibitem{matacz1993quantum}
\bibinfo{author}{Matacz, A.}, \bibinfo{author}{Davies, P.} \& \bibinfo{author}{Ottewill, A.~C.}
\newblock \bibinfo{title}{Quantum vacuum instability near rotating stars}.
\newblock \emph{\bibinfo{journal}{Physical Review D}} \textbf{\bibinfo{volume}{47}}, \bibinfo{pages}{1557} (\bibinfo{year}{1993}).

\bibitem{albrecht1994inflation}
\bibinfo{author}{Albrecht, A.}, \bibinfo{author}{Ferreira, P.}, \bibinfo{author}{Joyce, M.} \& \bibinfo{author}{Prokopec, T.}
\newblock \bibinfo{title}{Inflation and squeezed quantum states}.
\newblock \emph{\bibinfo{journal}{Physical Review D}} \textbf{\bibinfo{volume}{50}}, \bibinfo{pages}{4807} (\bibinfo{year}{1994}).

\bibitem{gasperini1993quantum}
\bibinfo{author}{Gasperini, M.} \& \bibinfo{author}{Giovannini, M.}
\newblock \bibinfo{title}{Quantum squeezing and cosmological entropy production}.
\newblock \emph{\bibinfo{journal}{Classical and Quantum Gravity}} \textbf{\bibinfo{volume}{10}}, \bibinfo{pages}{L133} (\bibinfo{year}{1993}).

\bibitem{hu1994squeezed}
\bibinfo{author}{Hu, B.~L.}, \bibinfo{author}{Kang, G.} \& \bibinfo{author}{Matacz, A.}
\newblock \bibinfo{title}{Squeezed vacua and the quantum statistics of cosmological particle creation}.
\newblock \emph{\bibinfo{journal}{International Journal of Modern Physics A}} \textbf{\bibinfo{volume}{9}}, \bibinfo{pages}{991--1007} (\bibinfo{year}{1994}).

\bibitem{nieto1997displaced}
\bibinfo{author}{Nieto, M.~M.}
\newblock \bibinfo{title}{Displaced and squeezed number states}.
\newblock \emph{\bibinfo{journal}{Physics Letters A}} \textbf{\bibinfo{volume}{229}}, \bibinfo{pages}{135--143} (\bibinfo{year}{1997}).

\bibitem{ellis1999deviation}
\bibinfo{author}{Ellis, G.~F.} \& \bibinfo{author}{Van~Elst, H.}
\newblock \bibinfo{title}{Deviation of geodesics in flrw spacetime geometries}.
\newblock \emph{\bibinfo{journal}{On Einstein’s Path: Essays in Honor of Engelbert Schucking}} \bibinfo{pages}{203--225} (\bibinfo{year}{1999}).

\bibitem{penzias1965measurement}
\bibinfo{author}{Penzias, A.~A.} \& \bibinfo{author}{Wilson, R.~W.}
\newblock \bibinfo{title}{A measurement of excess antenna temperature at 4080 mc/s.}
\newblock \emph{\bibinfo{journal}{Astrophysical Journal, vol. 142, p. 419-421}} \textbf{\bibinfo{volume}{142}}, \bibinfo{pages}{419--421} (\bibinfo{year}{1965}).

\bibitem{handley_curvature_2021}
\bibinfo{author}{Handley, W.}
\newblock \bibinfo{title}{Curvature tension: {Evidence} for a closed universe}.
\newblock \emph{\bibinfo{journal}{Physical Review D}} \textbf{\bibinfo{volume}{103}}, \bibinfo{pages}{L041301} (\bibinfo{year}{2021}).
\newblock \urlprefix\url{https://link.aps.org/doi/10.1103/PhysRevD.103.L041301}.

\bibitem{savage1986inhibition}
\bibinfo{author}{Savage, C.} \& \bibinfo{author}{Walls, D.}
\newblock \bibinfo{title}{Inhibition of tunneling in optical bistability by a squeezed vacuum}.
\newblock \emph{\bibinfo{journal}{Physical review letters}} \textbf{\bibinfo{volume}{57}}, \bibinfo{pages}{2164} (\bibinfo{year}{1986}).

\bibitem{albrecht1982cosmology}
\bibinfo{author}{Albrecht, A.} \& \bibinfo{author}{Steinhardt, P.~J.}
\newblock \bibinfo{title}{Cosmology for grand unified theories with radiatively induced symmetry breaking}.
\newblock \emph{\bibinfo{journal}{Physical Review Letters}} \textbf{\bibinfo{volume}{48}}, \bibinfo{pages}{1220} (\bibinfo{year}{1982}).

\end{thebibliography}

\end{document}